\pdfoutput=1
\documentclass[sigconf,acmthm=false]{acmart}
\setcopyright{none}
\renewcommand\footnotetextcopyrightpermission[1]{}

\usepackage{algorithm}
\usepackage{algpseudocode}
\usepackage{amsfonts}
\usepackage{float}
\usepackage{graphicx}
\usepackage{amsmath} 
\usepackage{wrapfig}
\usepackage{caption}
\usepackage{multirow}
\usepackage{url}
\usepackage{booktabs}
\usepackage{adjustbox}
\usepackage[normalem]{ulem}
\usepackage{subcaption}
\usepackage{amsthm}
\usepackage{thm-restate}
\usepackage{cancel}

\usepackage{xcolor}
\usepackage{natbib}
\theoremstyle{definition}
\definecolor{mustard}{HTML}{E1AD01}

\newtheorem{theorem}{Theorem}
\newtheorem{corollary}{Corollary}
\newtheorem{remark}{Remark}

\usepackage[dvipsnames]{xcolor}
\definecolor{darkgreen}{rgb}{0,0.5,0}

\begin{document}

\keywords{Hamilton-Jacobi reachability analysis, Formal verification}

\title{A Quantitative Approach to Reach–Avoid Reachability Analysis for Discrete-Time Nonlinear Control Systems}

\author{Ihab Tabbara}
\email{i.k.tabbara@wustl.edu}
\affiliation{%
  \department{Department of Computer Science}
  \institution{Washington University in St.~Louis}
  \city{St.~Louis}
  \state{MO}
  \country{USA}}

\author{Eliya Badr}
\email{egb11@mail.aub.edu}
\affiliation{%
  \institution{American University of Beirut}
  \city{Beirut}
  \country{Lebanon}}

\author{Hussein Sibai}
\email{sibai@wustl.edu}
\affiliation{%
  \department{Department of Computer Science}
  \institution{Washington University in St.~Louis}
  \city{St.~Louis}
  \state{MO}
  \country{USA}}

\begin{abstract}

A reach-avoid specification requires a dynamical system to reach a target set of states while avoiding a failure set. Reachability analysis partitions the state space into parts starting from which the specification can be satisfied by following some control signal and those from which it cannot despite the best control effort. In this paper, we present a quantitative approach for reachability analysis to analyze the satisfaction of avoid and reach-avoid specifications for  discrete-time nonlinear control systems. In such an approach, a value function assigns a value to each state that determines how robustly the system satisfies or  violates the specification starting from it and following the optimal control signal. However, computing such a value function is difficult. Instead, our approach constructs discrete abstractions on which it performs value iteration to compute sound bounds on the desired value function. The resulting lower bound enables a sound under-approximation of the reach-avoid set (RAS) and, in the avoid case, a sound over-approximation of the backward reachable set (BRS). When the control set is finite,
the upper bound additionally enables a sound over-approximation of the RAS and, in the avoid case, a sound under-approximation of the BRS.  Moreover, the obtained bounds provide quantitative guarantees on safety margins and, in discounted reach-avoid problems, on safe target-reaching times. When the control set is finite, they also bound failure-reaching times.  Our approach further refines the abstraction by splitting 
the grid cells that could not be classified as within or outside the BRS (or RAS) to improve the bounds.
We evaluate our approach on five  systems  with up to five-dimensional state spaces. 

\end{abstract}
\maketitle
\pagestyle{plain}

\section{Introduction}
Assuring specification satisfaction for nonlinear control systems is crucial, especially for 
those deployed in safety-critical settings, such as autonomous vehicles \citep{end_to_end_AD_survey_2024} and surgical robots \cite{autonomy_surgical_robots}. 
Hamilton-Jacobi (HJ) reachability analysis \cite{mitchell2005HJ,mitchell2005toolbox} is  a fundamental technique addressing the challenge of formal safety verification and control synthesis. Given a user-defined failure 
set of states $\mathcal{F}$ that the system must avoid reaching, HJ reachability analysis results in the corresponding backward reachable set (BRS), which is the set of states starting from which reaching  the failure set cannot be avoided under any feasible control input. HJ reachability analysis represents finite-horizon backward reachable sets as nonpositive level sets of value functions and computes corresponding policies that maximize safety. When also given a target set $\mathcal{T}$, HJ reachability analysis represents the reach-avoid set (RAS)  consisting of the states starting from which the system can reach $\mathcal{T}$ while avoiding $\mathcal{F}$ as the super-zero level set of another value function and computes a corresponding policy that achieves the specification optimally. 

Existing HJ reachability analysis  methods \cite{mitchell2005toolbox,stanfordaslhjreachability(we_used_this),bui2022optimizeddp} solve the HJ partial differential equation (PDE) numerically on a  grid that discretizes the continuous state space.
However, they generally do not explicitly account for discretization errors and consequently do not guarantee that the computed sets over- or under-approximate the BRS or 
the RAS. 

In this paper, we develop a framework for computing sound lower and upper bounds on the HJ reachability value functions for discrete-time systems,   which define sound over-\hspace{0pt} and under-approximations of the BRS in the avoid case and
of the RAS in the reach-avoid (RA) case. 
Our algorithm discretizes the state space into a finite grid over which it computes two corresponding value functions $\underline{V}_\gamma$ and $\overline{V}_\gamma$ which lower-bound and, when the control set is finite, upper-bound the HJ value function  $V_\gamma$ over the continuous state space. Such bounds enable the evaluation of the robust satisfaction or violation of avoid and reach-avoid specifications, which cannot be inferred from the mere binary classification of states as within or outside the BRS or RAS.

Our algorithm computes $\overline{V}_\gamma$ and $\underline{V}_\gamma$
using value iteration. For discounted reach-avoid problems,
we choose initializations that make the  iterates of  $\underline{V}_\gamma$ 
non-decreasing and those of $\overline{V}_\gamma$ non-increasing, while
preserving their respective bounds.
Thus, the algorithm can be stopped at any iteration and still return sound bounds on $V_\gamma$.

Finally, our algorithm automatically and locally refines the grid  by splitting 
cells whose computed bounds $\underline{V}_\gamma$ and $\overline{V}_\gamma$ do not determine whether the cells are contained in or are disjoint from the BRS or RAS. 
By iteratively refining these cells, our algorithm can obtain a more accurate over-approximation of the BRS or under-approximation of the  RAS. 
We validate our algorithm in  
five case studies with up to five-dimensional state spaces.

Our main contributions are summarized below:
\begin{itemize}
\item We present an algorithm for computing sound bounds on avoid and reach-avoid reachability value functions of discrete-time nonlinear control systems while accounting for state-space discretization errors. These bounds provide quantitative guarantees on target-reaching times, safety margins, and failure-reaching times.
\item In the discounted reach-avoid case, we show that the bounds are sound at every iteration of value iteration, not only at the fixed point, and that, under certain conditions, they converge to the value function on uniform grids.
\item We present a refinement procedure that splits the cells in the grid which could not be classified 
to obtain more accurate approximations of the BRS and the RAS. 
\end{itemize}

\section{Related Work}
\label{sec:related_work}

HJ reachability methods compute the BRS or RAS for continuous-time systems by solving HJ partial differential equations over discretized state spaces. 
Symbolic control synthesis approaches construct finite abstractions of systems with continuous state spaces and synthesize corresponding controllers that are then refined to concrete ones that satisfy the user-provided formal specifications.
Viability-theoretic methods compute viability kernels of user-provided constraint sets, which are the largest subsets of these sets from which at least one trajectory remains in the constraint set for all time. 
More recently, learning-based approaches have been proposed for approximating reachability value functions. 
Our work is most closely related to HJ reachability methods and abstraction-based control synthesis approaches. In contrast to these works, our approach computes sound upper and lower bounds on avoid and reach-avoid value functions, which quantify the optimal robustness of specification satisfaction or violation at every initial state.

\subsection{Hamilton-Jacobi reachability analysis}
Existing HJ reachability analysis methods  for continuous-time systems~\cite{mitchell2005toolbox,stanfordaslhjreachability(we_used_this),bui2022optimizeddp, fisac2015reach, mitchell2003overapproximating, margellos2011hamilton} discretize the state space using a Cartesian grid and apply level-set methods to solve a PDE over the grid. 
Such approaches produce approximate solutions of PDEs, which only become exact as the grid spacing approaches zero \cite{mitchell2001validating}.
They are not guaranteed to result in sound over- or under-approximations of the BRS or the RAS. 

Numerical methods for approximating viscosity solutions of first-order  Hamilton-Jacobi equations have been extensively studied~\cite{falcone2016numerical}.  The Crandall-Lions theorem \cite{crandall1984two} establishes that monotone  finite-difference schemes converge to the viscosity solution with explicit  error bounds of $O(\Delta t^{1/2})$ shrinking as $\Delta x, \Delta t \to 0$, 
while the Barles-Souganidis theorem \cite{barles1991convergence} guarantees  convergence  without an explicit rate. Additionally, semi-Lagrangian schemes achieve an error bound of $O(\Delta x / \Delta t)$  under the condition $\Delta x = o(\Delta t)$ \cite{falcone2016numerical}.

We do not build on these results to bound the value function, because they analyze PDE solvers used for analyzing continuous-time systems, whereas we consider discrete-time systems.

Eqtami and Girard~\cite{eqtami2018safety} compute the avoid value function of finite transition systems and, via approximate alternating simulations, certify safety margins of infinite transition systems. We additionally address reach-avoid specifications, provide lower and upper bounds at every discounted iteration, derive guarantees on failure-reaching times in the avoid setting, and adaptively refine unclassified cells.

\subsection{Symbolic control synthesis}

Symbolic control approaches solve optimal control problems by constructing finite abstractions of the system dynamics and synthesizing controllers over the resulting transition systems \cite{Discrete_abstractions_of_nonlinear_systems_based_on_error_propagation_analysis_2011_9insymbolicoptimalcontrol, Efficient_solution_of_optimal_control_problems_using_hybrid_systems_2005_8insymbolicoptimalcontrol, Optimal_control_with_regular_objectives_using_an_abstraction_refinement_approach_2016_10insymbolicoptimalcontrol, Symbolic_approximate_time_optimal_control_2011_7inState-feedback_Abstractions_for_Optimal_Control_of_Piecewise-affine_Systems, Controller_synthesis_for_safety_and_reachability_via_approximate_bisimulation_2012_8inState-feedback_Abstractions_for_Optimal_Control_of_Piecewise-affine_Systems, Symbolicoptimalcontrol_2018}. The abstract systems are related to the concrete ones through 
simulation relations.

For example, Tazaki and Imura~\cite{Discrete_abstractions_of_nonlinear_systems_based_on_error_propagation_analysis_2011_9insymbolicoptimalcontrol} construct discrete abstractions with approximate bisimulation guarantees and compute optimal control solutions on the abstractions. Leong and Prabhakar~\cite{Optimal_control_with_regular_objectives_using_an_abstraction_refinement_approach_2016_10insymbolicoptimalcontrol} construct weighted transition systems whose optimal costs provide bounds on the optimal control costs of the original systems. Mazo and Tabuada~\cite{Symbolic_approximate_time_optimal_control_2011_7inState-feedback_Abstractions_for_Optimal_Control_of_Piecewise-affine_Systems}
similarly construct abstract systems to compute approximately time-optimal controllers for the concrete ones, while Girard  \cite{Controller_synthesis_for_safety_and_reachability_via_approximate_bisimulation_2012_8inState-feedback_Abstractions_for_Optimal_Control_of_Piecewise-affine_Systems} proposes a method to synthesize controllers using approximately bisimilar abstractions for safety and reachability while providing bounds on the distance to the time-optimal controller. 
These approaches quantify optimality with respect to objectives such as reach time or accumulated cost. Our approach instead computes bounds on a reach-avoid value function that quantifies the degree of satisfaction or violation of the reach-avoid specification when the system starts  from different initial states.
Also, unlike our method, the approach in \cite{Efficient_solution_of_optimal_control_problems_using_hybrid_systems_2005_8insymbolicoptimalcontrol} relies on the ability to exactly determine first integrals of the plant dynamics, those of  \cite{Efficient_solution_of_optimal_control_problems_using_hybrid_systems_2005_8insymbolicoptimalcontrol} and \cite{Discrete_abstractions_of_nonlinear_systems_based_on_error_propagation_analysis_2011_9insymbolicoptimalcontrol} do not work  with hard constraints, and \cite{Optimal_control_with_regular_objectives_using_an_abstraction_refinement_approach_2016_10insymbolicoptimalcontrol} only applies to discrete-time piecewise-linear systems with continuous cost functions.

Reissig and Rungger~\cite{Symbolicoptimalcontrol_2018} propose a symbolic optimal control framework. They construct finite abstractions on covers of the state space and prove that their value functions upper-bound the true value function and converge to it as the abstraction’s conservatism, a parameter bounding its approximation errors, tends to zero. In their reach-avoid formulation, the value function equals zero on the RAS and infinity elsewhere, characterizing satisfaction without quantifying robustness. In contrast, our RA value function quantifies the robustness of satisfaction and violation of the RA specification. In their minimum-time reach-avoid formulation, the value function is the minimum time to reach the target safely, so their computed values upper-bound this time. In our work, for discounted RA problems, a positive lower bound yields a maximum target-reaching time and a minimum value of the failure function until then, and, if the control set is finite, a negative upper bound yields a maximum time within which every policy reaches the failure set (Proposition~\ref{prop:greedy}). In both of their formulations, the computed values do not generally certify that a state is outside the RAS. Our lower and upper bounds yield an under-approximation and, when the control set is finite, an over-approximation of the RAS.

Control synthesis tools such as CoSyMA~\cite{CoSyM_control_synthesismultiscale_abstractions_2013} and ROCS~\cite{rocs_2018,ROCS2} implement symbolic control
algorithms. ROCS~2.0, which we use as a baseline,
supports invariance, reachability, and LTL specifications recognizable by deterministic B\"uchi automata. It 
can compute sound under-approximations of winning sets, i.e., initial sets of states from which a controller can enforce a given specification. The methods underlying ROCS account for state and input discretization and establish robust completeness under the corresponding regularity, labeling, and resolution assumptions~\cite{liu2017robust, li2021robustly,li2022specification}. 

As in ROCS~2.0, we compute sound under-approximations of the winning set (the reach-avoid set), including with continuous control sets. When the control set is finite, our algorithm additionally computes an under-approximation of the set of states from which no admissible controller can reach the target safely, which ROCS~2.0 does not compute. A second difference is that our computed bounds enable quantitative evaluation of the robust satisfaction or violation of avoid and reach-avoid specifications, beyond mere binary classification as within or outside the BRS or RAS.

\subsection{Viability theory}

Viability theory \cite{saint1994approximation, cardaliaguet1999set} aims to compute the viability kernel, which is the largest set of states starting from which a dynamical system can stay within a constraint set indefinitely under appropriate control actions. 
When the constraint set represents the complement of a failure set, the viability kernel is the largest safe controlled invariant set, the complement of the BRS. Related methods also exist for pursuit-evasion and differential games  \cite{Fully-discrete_schemes_for_the_value_function_of_pursuit-evasion_games_with_state_constraints}.

Similar to the previously discussed methods, existing numerical algorithms for computing viability kernels typically rely on grid discretizations of the state space and provide guarantees only asymptotically as discretization parameters approach zero. For finite grid resolutions, the computed kernel may differ from the true viability kernel. In particular, as discussed in \cite[Theorem 2.19]{cardaliaguet1999set}, the computed kernel may neither over- nor under-approximate the true viability kernel for discrete-time systems, and particularly, it may classify unsafe states as safe. Instead, our method 
computes over-approximations of the BRS and under-approximations of the RAS for any grid resolution. Also, rather than performing binary classification, we compute value functions, enabling  quantified evaluation of  robust specification satisfaction or violation.

\subsection{Neural-network approximations of reach-avoid value functions}

Recent work has explored learning-based approaches for approximating reachability value functions using neural networks. 
For example, Li et al.~\cite{li2025certifiable} propose training neural networks using deep reinforcement learning to approximate discounted reach-avoid value functions and then certifying neighborhoods around query states. 
Other works have investigated approximating the HJ PDE solution \cite{deepreach, dgm} using physics-informed neural networks \cite{physicsinformedNN} 
and sinusoidal networks \cite{sinusoidal_networks}.
While these approaches aim to improve scalability and mitigate the curse of dimensionality, the resulting neural approximations generally 
are not guaranteed to be correct.

\subsection{Other related work}

Another related line of research considers relaxations of invariance-based safety specifications. Liu and Mallada \cite{liu2025recurrentcontrolbarrierfunctions} introduce Recurrent Control Barrier Functions (RCBF), which relax the strict invariance requirement of classical control barrier functions to finite-time recurrence. RCBFs enable sound and efficient trajectory-based synthesis of under-approximations of the safe set, i.e., the complement of the BRS, and equivalently, over-approximations of the BRS.

Zhao et al.~\cite{zhao2022outer}, Wu et al.~\cite{wu2026controlled},
and Zhao et al.~\cite{zhao2023robust} use sum-of-squares optimization to compute sound approximations of reach-avoid sets for discrete-time polynomial systems. The first work computes over-approximations,
while the latter two compute under-approximations of these sets. Xue et al.~\cite{xue2020inner} similarly compute sound under-approximations of reach-avoid sets, but consider autonomous systems without control inputs.
Unlike our approach, these works only consider polynomial dynamics and do not compute bounds on the reach-avoid value functions.

Our bounds are also related to  abstraction-based approaches~\cite{adnane_discretize_2024, cauchi2019efficiency, dutreix2021specification} and to interval iteration for finite MDPs~\cite{haddad2018interval}. Alaoui and Saoud~\cite{adnane_discretize_2024} bound the Q-function in Q-learning, rather than a reach-avoid value function, on a grid whose resolution is precomputed based on a user-specified error bound, while our algorithm starts from an arbitrary resolution and refines only unclassified cells. Cauchi et al.~\cite{cauchi2019efficiency} and Dutreix and Coogan~\cite{dutreix2021specification} bound the probability of satisfying temporal logic specifications for switched linear systems with Gaussian noise and for mixed-monotone systems without control inputs, respectively. Haddad and Monmege~\cite{haddad2018interval} compute lower and upper bounds on the reachability probabilities of finite MDPs at every iteration, after a graph preprocessing that makes the fixed points unique. In contrast, our bounds apply to deterministic nonlinear control systems with Lipschitz dynamics and continuous state spaces, and, in the discounted setting, our discrete Bellman operators are contractions, so their fixed points are unique. Moreover, unlike these works~\cite{adnane_discretize_2024, cauchi2019efficiency, dutreix2021specification, haddad2018interval}, we quantify how robustly reach-avoid specifications are satisfied or violated, with safety margins and target- and failure-reaching times.

\section{Preliminaries}
\label{sec:HJ_value_function}

We consider systems with deterministic  discrete-time dynamics: 
\begin{equation}
x_{t+1} \;=\; f(x_t,u_t),\label{eq:dynamics}
\end{equation}
where $x_t \in \mathbb{R}^n$ denotes the system's state at time $t$,  
$u_t \in \mathcal{U} \subseteq \mathbb{R}^m$ denotes its
control input at time $t$, 
$\mathcal{U}$ is a compact set, and $f:\mathbb{R}^n \times \mathcal{U} \rightarrow \mathbb{R}^n$ is a nonlinear function. Let $\mathcal X\subseteq\mathbb R^n$ be a compact set that defines the domain of interest. We assume that the restriction of $f$ to $\mathcal X\times\mathcal U$ is Lipschitz continuous in the first argument with Lipschitz constant $L_f$ with respect to the infinity norm, i.e., $\|f(x,u)-f(y,u)\|_\infty\le L_f\|x-y\|_\infty$ for all $x,y\in\mathcal X$ and $u\in\mathcal U$.
We denote the trajectory with initial state $x \in \mathbb{R}^n$ and following a policy $\pi: \mathbb{R}^n\to\mathcal{U}$ by 
$\xi_x^\pi := \{x_t\}_{t\ge 0}$.

We encode a user-defined failure set $\mathcal{F} \supseteq \mathbb{R}^n \setminus \mathcal{X}$ as the {\em sub-zero level set} of a bounded function $l:\mathbb{R}^n\to\mathbb{R}$, whose restriction to $\mathcal X$ is Lipschitz continuous with respect to the infinity norm with a known Lipschitz constant $L_l>0$, i.e., $\mathcal{F}:= \{ x \in \mathbb{R}^n \ |\ l(x) \leq 0 \}$. Similarly, we assume that there exists a bounded reward
function \( r : \mathbb{R}^n\to \mathbb{R} \), whose restriction to $\mathcal X$ is Lipschitz continuous with respect to the infinity norm with a known Lipschitz constant $L_r>0$, defining the target set \( \mathcal{T} \subseteq \mathcal{X} \) as follows: $\mathcal{T}:=\{x \in \mathbb{R}^n\ |\  r(x)>0\} $. We assume that the target set and the failure set are disjoint, i.e., $\mathcal{T}\cap\mathcal{F}=\emptyset$. 
 Our goal is to compute the set of states starting from which the system~(\ref{eq:dynamics}) can be controlled to reach the target set safely.
 We refer to this set as the \emph{reach-avoid set} (RAS)  $\mathcal{R}$ defined as follows:
\begin{align}
    \mathcal{R} := \{
        &x_0  : \exists \pi \ \text{s.t.} \, \ \exists T < \infty,
         x_T \in \mathcal{T} \ \wedge \ (\forall t \in [0,T], \nonumber 
        x_t \notin \mathcal{F})\}.   \label{eq:RA_set}
    \end{align}

 Next, we characterize $\mathcal{R}$ as the super-zero level set of a value function.
Fix a discount factor $\gamma\in(0,1]$  
and define the {\em reach-avoid (RA) measure} \( g_{\gamma}(\xi^{\pi}_{x_0}, t) \) as
\begin{equation}
    g_{\gamma}(\xi^{\pi}_{x_0}, t) :=
    \min \left(
        \gamma^{t} \, r(x_t), \;
        \min_{\tau = 0, \dots, t} \gamma^{\tau} \, l(x_{\tau})
    \right),
    \label{eq:RA_measure}
\end{equation}
which is positive
if and only if  the trajectory $\xi^{\pi}_{x_0}$
reaches the target set $\mathcal{T}$ safely at time $t$.
Intuitively, \( g_{\gamma} \) measures both the progress towards the target and safety. At time \( t \), it returns the minimum, and therefore the 
worst-case, discounted value among the target reward and all intermediate safety 
costs encountered along the trajectory. 
Next, we define the time-discounted RA value function: 
\begin{equation}V_{\gamma}(x) := \sup_{\pi} \sup_{t \geq 0}
    g_{\gamma}(\xi^{\pi}_{x}, t).
    \label{eq:Vgamma_def}
\end{equation}

The value function \(V_{\gamma}\) evaluates the best achievable discounted 
reach-avoid measure starting from state \(x\), as the supremum of the RA measure over all policies and time horizons.

For any $\gamma \in (0,1]$, the super-zero level set
$
  \mathcal{R_{\gamma}}:=\{ x \in\mathbb R^n \mid V_{\gamma}(x) > 0 \}
$
equals the RAS $\mathcal{R}$ \cite{li2025certifiable}. 
Conversely, its complement,
$
    \mathcal{L}_{\gamma} := \{ x\in\mathbb R^n \mid V_{\gamma}(x) \leq 0 \},
$
defines the set of states from 
which no admissible policy can guarantee  
reaching the target set safely in finite time.
We thus drop the $\gamma$ subscript from $\mathcal{L}_\gamma$ and  $\mathcal{R_{\gamma}}$ and call them $\mathcal{L}$  and $\mathcal{R}$ for the rest of the paper. 

The Bellman operator associated with $V_\gamma$ is
defined as follows: 
\begin{equation}\mathcal{B}_\gamma[W](x) :=
    \sup_{u \in \mathcal{U}}   \;
    \min\bigl\{\, l(x),\;
        \max\{\, r(x),\; \gamma W(f(x,u)) \,\}
    \bigr\}.
    \label{eq:bellman_operator}
\end{equation}

For
any $\gamma\in(0,1)$, this operator
  is a contraction mapping on bounded functions $W:\mathbb{R}^n\to\mathbb R$, and $V_\gamma$ is its
unique fixed point~\cite{li2025certifiable}.

The results established so far characterize the reach-avoid value function $V_\gamma$. Since $\mathcal F\supseteq\mathbb R^n\setminus\mathcal X$, \eqref{eq:RA_measure} and~\eqref{eq:Vgamma_def} imply that $V_\gamma(x)\le l(x)\le0$, i.e., $x\in\mathcal L$, for all $x\in\mathbb R^n\setminus\mathcal X$. Thus, we focus on computing $V_\gamma$ over $\mathcal{X}$.
However, computing $V_\gamma$ exactly for nonlinear systems is generally intractable. Existing approaches approximate it by considering only particular states in the cells of a grid over $\mathcal{X}$ during its computation \cite{mitchell2005toolbox,mitchell2005HJ}. Instead, we bound its values from both sides by computing two value
functions over the grid.

\section{Discrete Abstractions} 
\label{sec:discrete_abstractions}

To bound $V_\gamma$, we build a {\em discrete abstraction}  $\mathcal{A} = (S, A, \Delta)$ of system~(\ref{eq:dynamics})~\cite{tabuada2009verification}. The abstraction $\mathcal{A}$ has a finite set of states $S$  representing the cells in a grid $\mathcal{G}$ over $\mathcal{X}$ along with a special state $s_\bot$ representing all states not in $\mathcal{X}$, a finite set of controls $A$, and a non-deterministic transition relation $\Delta$ that maps state-action pairs to subsets of $S$. 
The set of control actions $A$ is a finite subset of $\mathcal{U}$. 
One can, for example, construct $A$ as the set of centers of the cells of a grid over $\mathcal{U}$. 
To compute the transition relation $\Delta$, we compute the reachable sets of system~(\ref{eq:dynamics}) starting from each cell as the set of initial states and following each of the discrete control actions. 

We denote the one time-step reachable set starting from the cell represented by $s \in S$ and following control action $a \in A$ by 
$R(s, a) := \bigcup_{x \in [[s]]} \{f(x, a)\}$, 
where $[[s]]$ represents the set of states in $\mathcal{X}$ that belong to the cell in $\mathcal{G}$ represented by $s$, for $s\neq s_\bot$, and $[[s_\bot]]:=\mathbb R^n\setminus\mathcal X$. 
We assume that the cells cover $\mathcal X$ and are contained in $\mathcal X$. Thus, the sets $[[s]]$, $s\in S$, cover $\mathbb R^n$, similar to the abstractions in~\cite{weber2017optimized}, whose states cover $\mathbb R^n$ by uniform grid cells and overflow symbols.

One can compute an over-approximation of $R(s,a)$ by simulating system~(\ref{eq:dynamics}) forward starting from $x_c(s)$, the center of the cell $s$,  following $a$ and then constructing a ball of radius $L_f \max_{i} \eta_{x,i}(s)$, where $\eta_x(s) \in (\mathbb{R}^+)^n$ represents the radii in every dimension of the cell $s$, i.e., half of the side length in every dimension. More formally, $R(s,a) \subseteq B_{L_f \max_{i \in [n]}\eta_{x,i}(s)}(f(x_c(s),a))$, where $B_r(x)$ is the ball of infinity norm radius $r$, i.e., a hypercube of sides equal to $2r$ centered at $x \in \mathbb{R}^n$. This follows from $L_f$ being the Lipschitz constant of $f$ on $\mathcal X$, since $[[s]]\subseteq\mathcal X$ and $x_c(s)\in\mathcal X$. If there are $\varepsilon_1,\ldots,\varepsilon_n>0$ such that $\eta_{x,i}(s)=\varepsilon_i$ for all $i$ and all $s\in S\setminus\{s_\bot\}$, we call $\mathcal G$ uniform with cell radius $\varepsilon=\max_i\varepsilon_i$.

Given an over-approximation of the reachable set, which we also denote by $R$, we can construct the non-deterministic transition relation $\Delta: S \times A \rightarrow 2^S$ as follows: $\forall s \in S\setminus\{s_\bot\}$, $\forall s'\in S$, $\forall a \in A$, 
\begin{align}
    s' \in \Delta(s,a) \text{ iff } [[s']] \cap R(s,a) \neq \emptyset.
\end{align}
As for $s_\bot$, we make it absorbing, i.e., $\Delta(s_\bot,a):=\{s_\bot\}$ for all $a$.

We define $\Delta(s,u)$ similarly for any action $u \in \mathcal{U}$, i.e., $u$ that does not necessarily belong to $A$. 

\section{Value Function Bounds}

\label{sec:valfunc_bounds}
{\emergencystretch=1em
To bound $V_\gamma$, we first construct upper and lower bounds on the values $l$ and $r$ can have at the states in  
each cell of the grid $\mathcal{G}$. Since the restriction of $l$ to $\mathcal X$ is $L_l$-Lipschitz, $\forall s \in S \setminus \{s_\bot\}$, $\forall x \in [[s]]$, 
$|l(x) - l(x_c(s))| \leq L_l\|x - x_c(s)\|_\infty \leq L_l \max_{i\in [n]}\eta_{x,i}(s)$.
Thus, for all $s \in S\setminus \{s_\bot\}$, we define
$\underline{l}(s) := l(x_c(s)) - L_l \max_{i\in[n]}\eta_{x,i}(s)$ and  
$\overline{l}(s) := l(x_c(s)) + L_l  \max_{i\in[n]}\eta_{x,i}(s)$ so that 
$\forall x \in [[s]]$, $\underline{l}(s) \leq l(x) \leq \overline{l}(s)$. 
Similarly, since the restriction of $r$ to $\mathcal X$ is $L_r$-Lipschitz, $\forall s \in S\setminus\{s_\bot\}$, $\forall x \in [[s]]$, 
$|r(x) - r(x_c(s))| \leq L_r\|x - x_c(s)\|_\infty \leq
L_r \max_{i\in [n]}\eta_{x,i}(s)$.
Thus, for all $s \in S\setminus\{s_\bot\}$, we define
$\underline{r}(s) := r(x_c(s)) - L_r \max_{i\in[n]}\eta_{x,i}(s)$ and  
$\overline{r}(s) := r(x_c(s)) + L_r  \max_{i\in[n]}\eta_{x,i}(s)$ so that 
$\forall x \in [[s]]$, $\underline{r}(s) \leq r(x) \leq \overline{r}(s)$. For $s_\bot$, we set $\underline l(s_\bot)=\underline r(s_\bot):=\underline c$ and $\overline l(s_\bot)=\overline r(s_\bot):=\overline c$, where the constants $\underline c$ and $\overline c$ satisfy $\underline c\le\min\{l(x),r(x)\}$ and $\max\{l(x),r(x)\}\le\overline c\le0$ for all $x\in\mathbb R^n\setminus\mathcal X$. Such constants exist because $l$ and $r$ are bounded, and $\overline c=0$ is always valid because $\mathcal F\supseteq\mathbb R^n\setminus\mathcal X$ and $\mathcal T\subseteq\mathcal X$ imply that $l(x)\le0$ and $r(x)\le0$ for all $x\in\mathbb R^n\setminus\mathcal X$. Since $\mathcal X$ is compact, $\mathbb R^n\setminus\mathcal X$ is nonempty, so $\underline c\le\overline c\le0$. Then, for all $s\in S$ and $x\in[[s]]$, $\underline l(s)\le l(x)\le\overline l(s)$ and $\underline r(s)\le r(x)\le\overline r(s)$.\par}

Having constructed the discrete abstraction $\mathcal{A}$ in Section~\ref{sec:discrete_abstractions}, and cell-wise bounds on $l$ and $r$, we now introduce new value functions defined over the discrete state space $S$ which we will use to compute upper and lower bounds on the value function $V_\gamma$ defined in (\ref{eq:Vgamma_def}) over the continuous state space.

First, we define maximal and minimal RA measures for discrete trajectories, analogous to the continuous RA measure in (\ref{eq:RA_measure}), and derive corresponding maximal and minimal  value functions. 
Next, for any $\gamma\in(0,1)$, we prove that their Bellman operators are contractions with unique fixed points.

For each discrete state $s \in S$, which represents a cell in the grid $\mathcal{G}$ or, for $s=s_\bot$, the set $\mathbb R^n\setminus\mathcal X$, we define upper and lower bounds on the RA measure 
over all states in $[[s]]$. Consider a discrete trajectory $\{s_{\tau}\}_{{\tau}=0}^t$ starting from cell $s_0$ and following a sequence of discrete actions $\{a_{\tau}\}_{{\tau}=0}^{t-1}$, where $a_{\tau} \in A$ for all ${\tau} = 0, \ldots, t-1$, and $s_{{\tau}+1} \in \Delta(s_{\tau}, a_{\tau})$ for all ${\tau} = 0, \ldots, t-1$.

\textbf{Upper bound RA measure:} The \emph{maximal} discounted RA measure for such a discrete trajectory at time $t$ is defined as
\begin{equation}
    \overline{g}_{\gamma}(\{s_{\tau}\}_{{\tau}=0}^t) :=
    \min \left(
        \gamma^{t} \, \overline{r}(s_t), \;
        \min_{\tau = 0, \dots, t} \gamma^{\tau} \, \overline{l}(s_{\tau})
    \right),
    \label{eq:discrete_RA_measure_upper}
\end{equation}

\textbf{Lower bound RA measure:} The \emph{minimal} discounted RA measure for the same trajectory is
\begin{equation}
    \underline{g}_{\gamma}(\{s_{\tau}\}_{{\tau}=0}^t) :=
    \min \left(
        \gamma^{t} \, \underline{r}(s_t), \;
        \min_{\tau = 0, \dots, t} \gamma^{\tau} \, \underline{l}(s_{\tau})
    \right),
    \label{eq:discrete_RA_measure_lower}
\end{equation}

To bound $V_\gamma$ in~\eqref{eq:Vgamma_def}, consider
 policies of the form $\pi:S\to A$. For each $\pi$, all
trajectory optimizations below are over infinite sequences
$\{s_\tau\}_{\tau\ge0}$ satisfying $s_0=s$ and
$s_{\tau+1}\in\Delta(s_\tau,\pi(s_\tau))$ for all
$\tau\ge0$. Define the \emph{maximal} and \emph{minimal}
discrete RA values as
\begin{align}
    \overline{V}_\gamma(s)
    &:= \max_\pi
    \max_{\{s_\tau\}_{\tau\ge0}}
    \sup_{t\ge0}
    \overline{g}_\gamma(\{s_\tau\}_{\tau=0}^{t}),
    \label{eq:discrete_value_upper}\\
    \underline{V}_\gamma(s)
    &:= \max_\pi
    \min_{\{s_\tau\}_{\tau\ge0}}
    \sup_{t\ge0}
    \underline{g}_\gamma(\{s_\tau\}_{\tau=0}^{t}).
    \label{eq:discrete_value_lower}
\end{align}
These values optimize feedback control under best-
and worst-case successor choices, respectively.
By an argument similar to that in~\cite{li2025certifiable}, they satisfy the Bellman equations:
\begin{align}
    \overline{V}_\gamma(s) = \max_{a \in A} \min\biggl\{\overline{l}(s), 
     \max\biggl\{\overline{r}(s), \max_{s' \in \Delta(s,a)} \gamma \overline{V}_\gamma(s')\biggr\}\biggr\}, \label{eq:bellman_upper_discrete} \\
    \underline{V}_\gamma(s) = \max_{a \in A} \min\biggl\{\underline{l}(s),  
     \max\biggl\{\underline{r}(s), \min_{s' \in \Delta(s,a)} \gamma \underline{V}_\gamma(s')\biggr\}\biggr\}. \label{eq:bellman_lower_discrete}
\end{align}
Both values depend on $\mathcal A=(S,A,\Delta)$. By Corollary~\ref{cor:restricted_bounds}, for $\gamma\in(0,1)$, $\underline V_\gamma$ is a lower bound on $V_\gamma$ for every nonempty and finite $A\subseteq\mathcal U$, but $\overline V_\gamma$ is only guaranteed to be an upper bound  if $A=\mathcal U$.

Define the discrete Bellman operators $\overline{\mathcal{T}}_\gamma$ and $\underline{\mathcal{T}}_\gamma$ as follows, for every $V:S\to\mathbb R$ and $s\in S$:
\begin{align*}
(\overline{\mathcal{T}}_\gamma V)(s) &:= \max_{a \in A} \min\{\overline{l}(s), \max\{\overline{r}(s), \max_{s' \in \Delta(s,a)} \gamma V(s')\}\}, \\
(\underline{\mathcal{T}}_\gamma V)(s) &:= \max_{a \in A} \min\{\underline{l}(s), \max\{\underline{r}(s), \min_{s' \in \Delta(s,a)} \gamma V(s')\}\}.
\end{align*}
Given initial functions $\underline V_\gamma^{(0)},\overline V_\gamma^{(0)}:S\to\mathbb R$, value iteration computes the iterates $\underline V_\gamma^{(k+1)}:=\underline{\mathcal T}_\gamma\underline V_\gamma^{(k)}$ and $\overline V_\gamma^{(k+1)}:=\overline{\mathcal T}_\gamma\overline V_\gamma^{(k)}$ for $k\geq0$.

\begin{restatable}[Contraction and Uniqueness]{proposition}{restateContraction}
\label{prop:discrete_contraction}
Fix a $\gamma \in (0,1)$. 
Then, for any two bounded functions $W_1, W_2 : S \rightarrow \mathbb{R}$,
\begin{align}
\|\overline{\mathcal{T}}_\gamma W_1 - \overline{\mathcal{T}}_\gamma W_2\|_\infty &\le \gamma\, \|W_1 - W_2\|_\infty, \label{eq:contraction_upper} \\
\|\underline{\mathcal{T}}_\gamma W_1 - \underline{\mathcal{T}}_\gamma W_2\|_\infty &\le \gamma\, \|W_1 - W_2\|_\infty. \label{eq:contraction_lower}
\end{align}
In particular, both operators are $\gamma$-contractions in the space of bounded functions on $S$ with the supremum norm, each admitting a unique fixed point $\overline{V}_\gamma$ and $\underline{V}_\gamma$, respectively.
\end{restatable}

\begin{proof}
The proof is given in Appendix~\ref{app:contraction}.
\end{proof}

Next, we show how to bound $V_\gamma$.
\begin{theorem}
\label{thm:soundness_monotone_bounds}
Fix $\gamma \in (0,1)$. Let  $\mathcal{A}$ be the abstraction constructed in
Section~\ref{sec:discrete_abstractions} and assume that $\mathcal{U}$ is finite and $A=\mathcal{U}$.
Then, initialize
\begin{align*}
\underline V_\gamma^{(0)}(s)
    = \min\{\underline l(s),\underline r(s)\} \text{ and }
\overline V_\gamma^{(0)}(s)
    = \overline l(s).
\end{align*}
Let $\underline V_\gamma$ and $\overline V_\gamma$ denote the
unique fixed points established in Proposition~\ref{prop:discrete_contraction}. Then, for every
$k\geq 0$, $s\in S$, and $x\in[[s]]$,
\begin{equation}
\label{eq:monotone-value-bounds}
\begin{aligned}
\underline V_\gamma^{(k)}(s)
&\leq \underline V_\gamma^{(k+1)}(s)
 \leq \underline V_\gamma(s)\\
&\leq V_\gamma(x)
 \leq \overline V_\gamma(s)
\leq \overline V_\gamma^{(k+1)}(s)
 \leq \overline V_\gamma^{(k)}(s).
\end{aligned}
\end{equation}
Consequently, value iteration can be stopped at any iteration and result in sound bounds. Finally, the sequences
$\{\underline V_\gamma^{(k)}\}_{k\geq0}$ and
$\{\overline V_\gamma^{(k)}\}_{k\geq0}$
converge uniformly on $S$ to $\underline V_\gamma$
and $\overline V_\gamma$, respectively.

\end{theorem}

\begin{proof}
We first establish monotonicity of the Bellman operators and
their properly initialized iterates. We then prove that every pair of
iterates bounds the concrete value function  $V_\gamma$. Finally, contraction
and taking the limit as the number of iterations increases to infinity establish the full inequality.

\noindent
\emph{Step 1: Monotonicity of the Bellman operators.}
Let $W,Z:S\to\mathbb R$ satisfy $W\leq Z$ pointwise.
Since $\gamma>0$, for every $s\in S$ and $a\in A$,
\begin{align*}
\min_{s'\in\Delta(s,a)}\gamma W(s')
&\leq
\min_{s'\in\Delta(s,a)}\gamma Z(s'),\\
\max_{s'\in\Delta(s,a)}\gamma W(s')
&\leq
\max_{s'\in\Delta(s,a)}\gamma Z(s').
\end{align*}
The remaining minimum and maximum operations preserve these
inequalities. Thus,
$\underline{\mathcal T}_\gamma W
 \leq\underline{\mathcal T}_\gamma Z$
and
$\overline{\mathcal T}_\gamma W
 \leq\overline{\mathcal T}_\gamma Z$.

\noindent
\emph{Step 2: Monotonicity of the iterate sequences.}
For every $a\in A$, \allowbreak
$\max\{\underline r(s),$ \allowbreak
$\min_{s'\in\Delta(s,a)}$ \allowbreak
$\gamma\underline V_\gamma^{(0)}(s')\}$ \allowbreak
$\geq\underline r(s)$.
Taking the minimum with $\underline l(s)$ and then
maximizing over $a\in A$ gives \allowbreak
$\underline V_\gamma^{(1)}(s)$ \allowbreak
$\geq\min\{\underline l(s),$ \allowbreak
$\underline r(s)\}$ \allowbreak
$=\underline V_\gamma^{(0)}(s)$.
Similarly, for every $a\in A$, \allowbreak
$\min\{\overline l(s),$ \allowbreak
$\max\{\overline r(s),$ \allowbreak
$\max_{s'\in\Delta(s,a)}$ \allowbreak
$\gamma\overline V_\gamma^{(0)}(s')\}\}$ \allowbreak
$\leq\overline l(s)$.
Maximizing over $a\in A$ gives \allowbreak
$\overline V_\gamma^{(1)}(s)$ \allowbreak
$\leq\overline l(s)$ \allowbreak
$=\overline V_\gamma^{(0)}(s)$.
These are the base cases for induction. Suppose that
$\underline V_\gamma^{(k)}
 \leq\underline V_\gamma^{(k+1)}$
and
$\overline V_\gamma^{(k+1)}
 \leq\overline V_\gamma^{(k)}$.
Step~1 yields
\begin{align*}
\underline{\mathcal T}_\gamma
\underline V_\gamma^{(k+1)}
&\geq
\underline{\mathcal T}_\gamma
\underline V_\gamma^{(k)},\\
\overline{\mathcal T}_\gamma
\overline V_\gamma^{(k+1)}
&\leq
\overline{\mathcal T}_\gamma
\overline V_\gamma^{(k)}.
\end{align*}
Thus,
$\underline V_\gamma^{(k+2)}
 \geq\underline V_\gamma^{(k+1)}$
and
$\overline V_\gamma^{(k+2)}
 \leq\overline V_\gamma^{(k+1)}$. 
 Hence, $\{\underline{V}_\gamma^{(k)}\}_{k \geq 0}$ is non-decreasing and $\{\overline{V}_\gamma^{(k)}\}_{k \geq 0}$ is non-increasing by induction.

\noindent
\emph{Step 3: Soundness of the iterates as bounds on $V_\gamma$.}
We prove by induction that
$\underline V_\gamma^{(k)}(s)
 \leq V_\gamma(x)
 \leq\overline V_\gamma^{(k)}(s)$
for every $s\in S$ and $x\in[[s]]$.
The initializations, the cell-wise bounds, and~\eqref{eq:Vgamma_def} give
\[
\begin{aligned}
\underline V_\gamma^{(0)}(s)
&= \min\{\underline l(s),\underline r(s)\}
\leq \min\{l(x),r(x)\}\\
&\leq V_\gamma(x)
\leq l(x)
\leq \overline l(s)
= \overline V_\gamma^{(0)}(s).
\end{aligned}
\]
Suppose the claim holds at iteration $k$ for all $s\in S$ and $x\in[[s]]$. For $s=s_\bot$, the claim holds at iteration $k+1$: the initialization and the definitions of $\underline{\mathcal T}_\gamma$ and $\overline{\mathcal T}_\gamma$ imply that $\underline V_\gamma^{(k)}(s_\bot)=\underline c$ and $\overline V_\gamma^{(k)}(s_\bot)=\overline c$ for every $k$, and every $x\in[[s_\bot]]$ satisfies $\underline c\le\min\{l(x),r(x)\}\le V_\gamma(x)\le l(x)\le\overline c$ by~\eqref{eq:RA_measure} and~\eqref{eq:Vgamma_def}. Fix $s\in S\setminus\{s_\bot\}$, $x\in[[s]]$, and
$a\in A$. By the construction of the abstraction in
Section~\ref{sec:discrete_abstractions}, there exists
$s_a\in\Delta(s,a)$ such that $f(x,a)\in[[s_a]]$, where $s_a=s_\bot$ if $f(x,a)\notin\mathcal X$. 
The induction hypothesis at this successor gives

\begin{align*}
\min_{s'\in\Delta(s,a)}
\underline V_\gamma^{(k)}(s')
&\leq \underline V_\gamma^{(k)}(s_a)
\leq V_\gamma(f(x,a))  \text{ and } \\
V_\gamma(f(x,a))
\leq \overline V_\gamma^{(k)}(s_a)
&\leq
\max_{s'\in\Delta(s,a)}
\overline V_\gamma^{(k)}(s').
\end{align*}
Combining these inequalities with the fact that $\gamma>0$, the cell-wise
bounds on $l$ and $r$, the monotonicity of minimum and maximum operators,
and the maximization over $a\in A=\mathcal U$ yields
\[
\underline V_\gamma^{(k+1)}(s)
\leq \mathcal B_\gamma[V_\gamma](x)
= V_\gamma(x)
\leq \overline V_\gamma^{(k+1)}(s).
\]
This completes the induction.

\noindent
\emph{Step 4: Convergence and the full inequality chain.}
By Proposition~\ref{prop:discrete_contraction}, both abstract Bellman operators are
$\gamma$-contractions. Consequently, their iterates converge
uniformly to $\underline V_\gamma$ and $\overline V_\gamma$,
respectively. Letting $k\to\infty$ in Step~3 results in
\[
\underline V_\gamma(s)
\leq V_\gamma(x)
\leq\overline V_\gamma(s).
\]
Combining the monotonicity in Step~2, the convergence
$\underline V_\gamma^{(k)}\to\underline V_\gamma$ and
$\overline V_\gamma^{(k)}\to\overline V_\gamma$,
and the bounds
$\underline V_\gamma(s)\leq V_\gamma(x)
\leq\overline V_\gamma(s)$ established above
yields~\eqref{eq:monotone-value-bounds}.
\end{proof}

\begin{corollary}
\label{cor:restricted_bounds}
Fix $\gamma\in(0,1)$. Let  $\mathcal{A}$ be the abstraction constructed in
Section~\ref{sec:discrete_abstractions},  $A\subseteq\mathcal U$ be nonempty and finite, 
and initialize
$\underline V_\gamma^{(0)}$ and
$\overline V_\gamma^{(0)}$ as in Theorem \ref{thm:soundness_monotone_bounds}.
The sequence $\{\underline V_\gamma^{(k)}\}_{k\geq0}$
is non-decreasing and converges uniformly on $S$ to
$\underline V_\gamma$.
The sequence $\{\overline V_\gamma^{(k)}\}_{k\geq0}$
is non-increasing and converges uniformly on $S$ to
$\overline V_\gamma$. For every $k\geq0$, $s\in S$, and
$x\in[[s]]$,
\[
\underline V_\gamma^{(k)}(s)
\leq \underline V_\gamma(s)
\leq V_\gamma(x) \text{ and }
\overline V_\gamma(s)
\leq \overline V_\gamma^{(k)}(s).
\]
If $A=\mathcal U$, then additionally
$V_\gamma(x)\leq\overline V_\gamma(s)$.
\end{corollary}

\begin{proof}
Let $V_\gamma^A$ denote the value function~\eqref{eq:Vgamma_def} with the supremum taken over policies $\pi:\mathbb R^n\to A$. Applying Theorem~\ref{thm:soundness_monotone_bounds} to the
system with control set $A$ shows that
$\{\underline V_\gamma^{(k)}\}_{k\geq0}$ is non-decreasing
and converges uniformly to $\underline V_\gamma$, while
$\{\overline V_\gamma^{(k)}\}_{k\geq0}$ is non-increasing
and converges uniformly to $\overline V_\gamma$. 
Moreover, for every $s\in S$ and $x\in[[s]]$,
\[
\underline V_\gamma(s)
\leq V_\gamma^A(x)
\leq \overline V_\gamma(s).
\]
Since $A\subseteq\mathcal U$, every policy admissible under $A$
is also admissible under $\mathcal U$. Hence, $V_\gamma^A(x)\leq V_\gamma(x)$, and therefore
$\underline V_\gamma(s)\leq V_\gamma(x)$.
When $A=\mathcal U$, we have $V_\gamma^A=V_\gamma$,
giving $V_\gamma(x)\leq\overline V_\gamma(s)$.
\end{proof}

\section{Value Iteration and Soundness Guarantees}
Algorithm~\ref{alg:alg1} applies value iteration on a
given grid, and its Bellman
operators maximize over $A$. For $k\ge1$, define
\begin{align*}
\underline\delta_k
&:=\|\underline V_\gamma^{(k)}
-\underline V_\gamma^{(k-1)}\|_\infty,\\
\overline\delta_k
&:=\|\overline V_\gamma^{(k)}
-\overline V_\gamma^{(k-1)}\|_\infty.
\end{align*}
The algorithm halts when
$\underline\delta_k\le\underline\delta^*$ and
$\overline\delta_k\le\overline\delta^*$, for user-specified
tolerances $\underline\delta^*,\overline\delta^*>0$.
It returns the last iterates:
\[
\underline V_\gamma^{\mathrm{cons}}
:=\underline V_\gamma^{(k)},\qquad
\overline V_\gamma^{\mathrm{cons}}
:=\overline V_\gamma^{(k)}.
\]
Corollary \ref{cor:restricted_bounds} shows that 
$\underline V_\gamma^{\mathrm{cons}}$ and $\overline V_\gamma^{\mathrm{cons}}$
remain below $\underline V_\gamma$ and above
$\overline V_\gamma$, respectively. Hence, the
classification guarantees of Theorem~\ref{thm:soundness}
hold at every iteration. Smaller tolerances can tighten the bounds
at the cost of additional iterations.

\begin{algorithm}\caption{Computing sound bounds on $V_\gamma$
}
\label{alg:alg1}

\begin{algorithmic}[1]
\State \textbf{Input:} finite grid $\mathcal G$ over
$\mathcal X$, abstract state set $S$ including $s_\bot$, nonempty finite action
set $A\subseteq\mathcal U$, dynamics $f$, functions $l,r$, constants $\underline c,\overline c$,
Lipschitz constants $L_f,L_l,L_r$ (w.r.t.\ $\|\cdot\|_\infty$), discount factor
$\gamma\in(0,1)$,
$\underline\delta^*,\overline\delta^*>0$
\For{$s\in S$}
    \State Compute $\underline l(s),\overline l(s),
    \underline r(s),\overline r(s)$ as in
    Section~\ref{sec:valfunc_bounds}
    \For{$a\in A$}
        \State Compute $\Delta(s,a)$ as in Section~\ref{sec:discrete_abstractions}
        \label{ln:computing_Delta}
    \EndFor
    \State $\underline V_\gamma^{(0)}(s)
    \gets\min\{\underline l(s),\underline r(s)\}$,  $\overline V_\gamma^{(0)}(s)
    \gets\overline l(s)$
\EndFor
\State $k\gets0$
\Repeat
    \State $k\gets k+1$
    \State $\underline V_\gamma^{(k)}
    \gets\underline{\mathcal T}_\gamma
    (\underline V_\gamma^{(k-1)})$,
    $\overline V_\gamma^{(k)}
    \gets\overline{\mathcal T}_\gamma
    (\overline V_\gamma^{(k-1)})$
    \State $\underline\delta_k\gets
    \|\underline V_\gamma^{(k)}
    -\underline V_\gamma^{(k-1)}\|_\infty$,
    $\overline\delta_k\gets
    \|\overline V_\gamma^{(k)}
    -\overline V_\gamma^{(k-1)}\|_\infty$
\Until{$\underline\delta_k\le\underline\delta^*$
\textbf{and} $\overline\delta_k\le\overline\delta^*$}
\label{ln:whileloop}
\State $\underline V_\gamma^{\mathrm{cons}}
\gets\underline V_\gamma^{(k)}$,
$\overline V_\gamma^{\mathrm{cons}}
\gets\overline V_\gamma^{(k)}$
\State \Return $\underline V_\gamma^{\mathrm{cons}},
\overline V_\gamma^{\mathrm{cons}}$
\end{algorithmic}
\end{algorithm}

\begin{theorem}\label{thm:soundness}
Fix a $\gamma\in(0,1)$ and let $\underline{V}_\gamma^{\mathrm{cons}}, \overline{V}_\gamma^{\mathrm{cons}} : S \to \mathbb{R}$ be the value functions computed by Algorithm~\ref{alg:alg1}.  Then, for any $s \in S$,
\begin{equation}
\label{eq:soundness_safe}
\underline{V}^{\mathrm{cons}}_{\gamma}(s) > 0 \implies [[s]] \cap \mathcal{L} = \emptyset \text{~and~} [[s]] \subseteq \mathcal{R}.
\end{equation}
Moreover, if $\mathcal{U}$ is finite and $A = \mathcal{U}$, then 
\begin{equation}
\label{eq:soundness_unsafe}
\overline{V}^{\mathrm{cons}}_{\gamma}(s) \leq 0 \implies [[s]] \subseteq \mathcal{L}.
\end{equation}
\end{theorem}

\begin{proof}
Let $k$ be the iteration at which Algorithm~\ref{alg:alg1}
terminates. Its outputs are
$\underline V_\gamma^{\mathrm{cons}}
 =\underline V_\gamma^{(k)}$ and
$\overline V_\gamma^{\mathrm{cons}}
 =\overline V_\gamma^{(k)}$.
Fix $s\in S$. By Corollary~\ref{cor:restricted_bounds},  for every $x\in[[s]]$,
\[
\underline V_\gamma^{\mathrm{cons}}(s)
\leq V_\gamma(x).
\]
Thus, if $\underline V_\gamma^{\mathrm{cons}}(s)>0$, then
$V_\gamma(x)>0$ for every $x\in[[s]]$.
Since $\mathcal R=\{x\in\mathbb R^n:V_\gamma(x)>0\}$ and
$\mathcal L=\mathbb R^n\setminus\mathcal R$,
$[[s]]\subseteq\mathcal R$ and
$[[s]]\cap\mathcal L=\emptyset$.

Now assume that $\mathcal U$ is finite and $A=\mathcal U$.
Corollary~\ref{cor:restricted_bounds} implies that for every $x\in[[s]]$,
\[
V_\gamma(x)
\leq\overline V_\gamma^{\mathrm{cons}}(s).
\]
If $\overline V_\gamma^{\mathrm{cons}}(s)\leq0$, then
$V_\gamma(x)\leq0$ for every $x\in[[s]]$.
By the definition of $\mathcal L$, it follows that
$[[s]]\subseteq\mathcal L$.
\end{proof}

If $f$ is Lipschitz continuous in $u$, then the assumption $A=\mathcal U$ in~\eqref{eq:soundness_unsafe} can be dropped by computing the upper bound on another discrete abstraction whose transitions account for all controls in $\mathcal U$, not only those in $A$, as shown next.
\begin{restatable}{proposition}{restateControlCells}
    \label{prop:control_cells}
    Fix a $\gamma\in(0,1)$. Assume that $f$ is Lipschitz continuous with respect to the second argument with Lipschitz constant $L_{f,u}$ with respect to the infinity norm, i.e.,  $\|f(x,u)-f(x,v)\|_\infty\leq L_{f,u}\|u-v\|_\infty$ for all $x\in\mathcal X$ and $u,v\in\mathcal U$, and let $A\subseteq\mathcal U$ be the set of centers of the cells of a grid whose cells cover $\mathcal U$, where $\eta_u(a)\in(\mathbb R^{\geq0})^m$ denotes the radii of the cell centered at $a\in A$. Let $\bar\Delta$ be constructed as $\Delta$ in Section~\ref{sec:discrete_abstractions} from the over-approximation $\bar R(s,a):=B_{L_f\max_i\eta_{x,i}(s)+L_{f,u}\max_j\eta_{u,j}(a)}(f(x_c(s),a))$. If $\overline{\mathcal T}_\gamma$ in Algorithm~\ref{alg:alg1} uses $\bar\Delta$ instead of $\Delta$, then $V_\gamma(x)\leq\overline V_\gamma^{(k)}(s)$ for all $k\geq0$, $s\in S$, and $x\in[[s]]$, and $\overline V_\gamma^{\mathrm{cons}}(s)\leq0$ implies $[[s]]\subseteq\mathcal L$.
    \end{restatable}
    \begin{proof}
The proof is given in Appendix~\ref{app:control_cells}.
    \end{proof}

In the remainder of the paper, we compute both bounds on $\Delta$. Hence, certificates of $\mathcal L$ based on the upper bound assume that $\mathcal U$ is finite and $A=\mathcal U$. That is also the case in our experiments. 
We next establish the same guarantees as provided in Theorem~\ref{thm:soundness} for the case when $\gamma=1$.

\begin{restatable}{proposition}{restateGammaOne}
\label{prop:gamma_1}
Let $\gamma=1$. In Algorithm~\ref{alg:alg1}, initialize $\underline V_1^{(0)}(s)=\min\{\underline l(s),\underline r(s)\}$ and $\overline V_1^{(0)}(s)=\min\{\overline l(s),\overline r(s)\}$ for every $s\in S$, and stop the value iteration when $\underline\delta_k=\overline\delta_k=0$. Then, Algorithm~\ref{alg:alg1} terminates after finitely many iterations, and its outputs $\underline V_1^{\mathrm{cons}}$ and $\overline V_1^{\mathrm{cons}}$ are fixed points of $\underline{\mathcal T}_1$ and $\overline{\mathcal T}_1$, respectively. Moreover, $\{\underline V_1^{(k)}\}_{k\geq0}$ is non-decreasing, and, for every $k\geq0$, $s\in S$, and $x\in[[s]]$, $\underline V_1^{(k)}(s)\le V_1(x)$. If $\mathcal U$ is finite and $A=\mathcal U$, then also $V_1(x)\le\overline V_1^{\mathrm{cons}}(s)$. 
For every $s\in S$, $\underline V_1^{\mathrm{cons}}(s)>0$ implies $[[s]]\subseteq\mathcal R$, and, if $\mathcal U$ is finite and $A=\mathcal U$, $\overline V_1^{\mathrm{cons}}(s)\le0$ implies $[[s]]\subseteq\mathcal L$.
\end{restatable}

\begin{proof}
The proof is given in Appendix~\ref{app:gamma_1}.
\end{proof}

Next, we derive quantitative guarantees for the satisfaction and the violation of the RA specification.

\begin{restatable}{proposition}{restateGreedy}
\label{prop:greedy}
Fix a $\gamma\in(0,1)$, let $\underline V_\gamma^{\mathrm{cons}}$ and $\overline V_\gamma^{\mathrm{cons}}$ be as returned by Algorithm~\ref{alg:alg1}, and fix some $x_0\in\mathcal X$ and let $s_0\in S$ be a state that satisfies  $x_0\in[[s_0]]$. Let $\hat\pi(s)\in\arg\max_{a\in A}\min_{s'\in\Delta(s,a)}\underline V_\gamma^{\mathrm{cons}}(s')$, $\forall s \in S$.
If $\underline V_\gamma^{\mathrm{cons}}(s_0)>0$, then the trajectory of system~\eqref{eq:dynamics} starting from $x_0$ and following $u_\tau=\hat\pi(s_\tau)$, where $s_\tau\in S$ satisfies $x_\tau\in[[s_\tau]]$, reaches $\mathcal T$ at some time $t\leq\lfloor\log(\underline V_\gamma^{\mathrm{cons}}(s_0)/r^*)/\log(\gamma)\rfloor$, where $r^*=\max_{s\in S}\underline r(s)$, and satisfies $l(x_\tau)\geq\gamma^{-\tau}\underline V_\gamma^{\mathrm{cons}}(s_0)>0$ for $\tau=0,\ldots,t$. If $\mathcal U$ is finite, $A=\mathcal U$, and $\overline V_\gamma^{\mathrm{cons}}(s_0)<0$, then, under any policy $\pi:\mathbb R^n\to\mathcal U$, the trajectory starting from $x_0$ reaches $\mathcal F$ at some time $t\leq\lfloor\log(\overline V_\gamma^{\mathrm{cons}}(s_0)/l_{\min})/\log(\gamma)\rfloor$, where $l_{\min}=\inf_{x\in\mathbb R^n}l(x)$. At this time, $l(x_t)\leq\gamma^{-t}\overline V_\gamma^{\mathrm{cons}}(s_0)<0$, and before this time, the trajectory does not reach $\mathcal T$, violating the RA specification.
\end{restatable}
\begin{proof}
The proof is given in Appendix~\ref{app:greedy}.
\end{proof}

If $f$ maps $\mathcal X\times\mathcal U$ into $\mathcal X$, i.e., trajectories starting in $\mathcal X$ remain in $\mathcal X$, then, on a uniform grid, the bounds computed by Algorithm~\ref{alg:alg1} become exact as the resolution of the grid becomes finer and the tolerances $\underline\delta^*$ and $\overline\delta^*$ approach zero, provided that $V_\gamma$ is Lipschitz continuous on $\mathcal X$, which holds when $\gamma L_f<1$~\cite{li2025certifiable}, as shown next.

\begin{restatable}{theorem}{restateConvergence}
\label{thm:convergence}
Assume that $f(x,u)\in\mathcal X$ for all $x\in\mathcal X$ and $u\in\mathcal U$. Fix a $\gamma\in(0,1)$ such that $\gamma L_f<1$, let $L:=\max\{L_l,L_r\}$, let $\mathcal U$ be finite and $A=\mathcal U$, let $\mathcal G$ be uniform with cell radius $\varepsilon$, and let $\Delta$ be constructed as in
Section~\ref{sec:discrete_abstractions}
with $R(s,a)=B_{L_f\varepsilon}(f(x_c(s),a))\cap\mathcal X$
for all $s\in S\setminus\{s_\bot\}$ and $a\in A$. Then, for every $s\in S\setminus\{s_\bot\}$ and $x\in[[s]]$,
\[
|\overline V_\gamma(s)-V_\gamma(x)|\leq\frac{2L\varepsilon}{1-\gamma} \text{ and }|\underline V_\gamma(s)-V_\gamma(x)|\leq\frac{2L\varepsilon}{1-\gamma}.
\]
Also, for every $s\in S\setminus\{s_\bot\}$ and $x\in[[s]]$, the outputs of Algorithm~\ref{alg:alg1} satisfy
\[
|\overline V_\gamma^{\mathrm{cons}}(s)-V_\gamma(x)|\leq\frac{2L\varepsilon+\gamma\overline\delta^*}{1-\gamma},\quad|\underline V_\gamma^{\mathrm{cons}}(s)-V_\gamma(x)|\leq\frac{2L\varepsilon+\gamma\underline\delta^*}{1-\gamma}.
\]
In particular, every cell $s$ left unclassified, i.e., with $\underline V_\gamma^{\mathrm{cons}}(s)\leq0<\overline V_\gamma^{\mathrm{cons}}(s)$, satisfies $|V_\gamma(x)|\leq(2L\varepsilon+\gamma\max\{\underline\delta^*,\overline\delta^*\})/(1-\gamma)$ for all $x\in[[s]]$.
\end{restatable}
\begin{proof}
The proof is given in Appendix~\ref{app:convergence}.
\end{proof}

After Algorithm \ref{alg:alg1} terminates, for any cell $s$ for which  $\underline{V}^{\mathrm{cons}}_{\gamma}(s) \leq 0 < \overline{V}^{\mathrm{cons}}_{\gamma}(s)$ or where  $\underline{V}^{\mathrm{cons}}_{\gamma}(s) \leq 0$ and $A\subset\mathcal U$, Theorem \ref{thm:soundness} does not determine 
$[[s]] \subseteq \mathcal{R} \text{~or~} [[s]] \subseteq \mathcal{L}$.

\section{Refining Unclassified Cells} 
\label{sec:refining_boundary}                        
Since a coarse grid may leave many cells unclassified, we present a refinement algorithm that splits them to better approximate $\mathcal L$ and $\mathcal R$. When $A\subset\mathcal U$, refinement can certify additional cells as belonging
to $\mathcal R$, but the computed upper bounds do not generally certify membership in $\mathcal L$. Upon termination of Algorithm~\ref{alg:alg1}, the cells in grid $\mathcal{G}$ are classified into three classes:

\begin{itemize}
    \item \emph{Cells in $\mathcal{R}$}: $\{ s \in S\ |\ \underline{V}_\gamma^{\mathrm{cons}}(s) > 0\}$,
    \item \emph{Cells in $\mathcal{L}$}: $\{s \in S\ |\ \mathcal{U}$ is finite, $A = \mathcal{U}$,  $\overline{V}_\gamma^{\mathrm{cons}}(s) \leq 0\}$,
    \item \emph{Unclassified cells}: any cell not in the previous two sets.
\end{itemize}

Algorithm~\ref{alg:alg2} implements a refinement scheme that, in each iteration, splits all queued cells along their longest sides into two equally-sized cells. The cell-wise bounds on the failure and reward functions, the transition relation $\Delta$, and the value function bounds are then recomputed via Algorithm~\ref{alg:alg1} on the refined grid. A cell is eligible for further refinement only when
its maximum radius exceeds the user-defined threshold $\varepsilon_{\min}$. This process is repeated until no cells satisfy the refinement condition. After the termination of
Algorithm~\ref{alg:alg2}, any remaining unclassified cells can be conservatively classified as belonging to $\mathcal{L}$, yielding an over-approximation of $\mathcal{L}$. Appendix~\ref{app:termination} discusses the cells that the queue omits when $A\subset\mathcal U$. 

\begin{algorithm}[t]
\caption{Adaptive refinement of unclassified cells}
\label{alg:alg2}
\begin{algorithmic}[1]
\State \textbf{Input:} $\mathcal{G}$, $\varepsilon_{\min}>0$, finite action set $A\subseteq\mathcal{U}$
\State Run Algorithm~\ref{alg:alg1} on $\mathcal{G}$ to initialize $\underline V_\gamma^{\mathrm{cons}},\overline V_\gamma^{\mathrm{cons}}$
\State $\mathit{queue} \gets \{s \in S\setminus\{s_\bot\}\ |\ \overline V_\gamma^{\mathrm{cons}}(s) > 0 \ \land \ \underline V_\gamma^{\mathrm{cons}}(s) \leq 0 \ \land \ \max_j \eta_{x,j}(s) > \varepsilon_{\min}\}$
\While{$\mathit{queue} \neq \emptyset$}
    \ForAll{$s \in \mathit{queue}$}
        \State Let $j^* = \arg\max_j \eta_{x,j}(s)$
        \State Split $s$ into two cells $s_1$ and $s_2$ along the $j^*$th axis
        \label{alg2:ln:split}
    \EndFor
    \State Recompute $\overline V_\gamma^{\mathrm{cons}}$ and $\underline V_\gamma^{\mathrm{cons}}$ over the new grid
    \label{alg2:ln:recompute_Vs}
    \State $\mathit{queue} \gets \{s \in S\setminus\{s_\bot\}\ |\ \overline V_\gamma^{\mathrm{cons}}(s) > 0 \ \land \ \underline V_\gamma^{\mathrm{cons}}(s) \leq 0 \ \land \ \max_j \eta_{x,j}(s) > \varepsilon_{\min}\}$
\EndWhile
\State \Return $\underline V_\gamma^{\mathrm{cons}},\overline V_\gamma^{\mathrm{cons}}$
\end{algorithmic}
\end{algorithm}

\label{sec:conservative_stopping}

\begin{restatable}{proposition}{restateTermination}\label{prop:termination}
Algorithms~\ref{alg:alg1} and~\ref{alg:alg2} terminate in finite time.
\end{restatable}

\begin{proof}
The proof is given in Appendix~\ref{app:termination}.
\end{proof}

\begin{remark}
    Extending our framework to continuous-time systems requires forward reachable set computations for continuous-time dynamics, a minimum over the continuous interval $[0,t]$ in $g_\gamma$ in~\eqref{eq:RA_measure}, and a supremum over admissible continuous-time control signals in~\eqref{eq:Vgamma_def}, \eqref{eq:discrete_value_upper}, and~\eqref{eq:discrete_value_lower}. We leave this extension to future work.
\end{remark}

\section{Avoid Value Function}
\label{sec:avoid value function}

In this section, $V_\gamma$ denotes the avoid value function rather than the reach-avoid value function~\eqref{eq:Vgamma_def}, and $\underline V_1^{(k)}$, $\overline V_1^{(k)}$, $\underline V_1^{\mathrm{cons}}$, and $\overline V_1^{\mathrm{cons}}$ denote the avoid quantities defined below. The avoid value function and Bellman operator are:
\begin{align}
V_\gamma(x)
&:= \sup_\pi \inf_{t \geq 0}
\min_{\tau=0,\ldots,t} \gamma^\tau l(x_\tau),
\label{eq:avoid_discounted_value_function} \\
\mathcal{B}_\gamma^{\mathrm{av}}[V](x)
&:= \min\{l(x), \sup_{u \in \mathcal{U}} \gamma V(f(x,u))\}.
\label{eq:avoid_bellman_operator}
\end{align}

For any state $x$ in the safe set, i.e., the complement of the BRS, a discount factor $\gamma\in(0,1)$ gives $V_\gamma(x)=0$. This is because $l$ is bounded, so $\gamma^\tau l(x_\tau)\to0$ along every trajectory. Hence, the infimum in~\eqref{eq:avoid_discounted_value_function} is at most zero, with equality along trajectories that avoid $\mathcal F$. Consequently, no lower bound on $V_\gamma$ can certify such
states using the test
$\underline V_\gamma^{\mathrm{cons}}(s)>0$.
We therefore set $\gamma=1$ in the avoid case to certify states that satisfy the avoid specification, and run value iteration until convergence. The abstract counterparts of~\eqref{eq:avoid_bellman_operator} for $\gamma=1$ are defined, for $V:S\to\mathbb R$ and $s\in S$, by
\begin{align*}
(\underline{\mathcal T}^{\mathrm{av}}V)(s)&:=\min\Bigl\{\underline l(s),\max_{a\in A}\min_{s'\in\Delta(s,a)}V(s')\Bigr\},\\
(\overline{\mathcal T}^{\mathrm{av}}V)(s)&:=\min\Bigl\{\overline l(s),\max_{a\in A}\max_{s'\in\Delta(s,a)}V(s')\Bigr\}.
\end{align*}
We initialize $\underline V_1^{(0)}:=\underline l$ and $\overline V_1^{(0)}:=\overline l$, and compute $\underline V_1^{(k+1)}:=\underline{\mathcal T}^{\mathrm{av}}\underline V_1^{(k)}$ and $\overline V_1^{(k+1)}:=\overline{\mathcal T}^{\mathrm{av}}\overline V_1^{(k)}$ for $k\geq0$. In the avoid case, we additionally require $\underline c\le\inf_{x\in\mathbb R^n}l(x)$, because the abstraction makes $s_\bot$ absorbing, whereas a trajectory of system~\eqref{eq:dynamics} that leaves $\mathcal X$ may return to $\mathcal X$ and then reach states where $l$ is smaller than anywhere outside $\mathcal X$.
The first update is less than or equal to the initialization
because $\min\{\underline l(s),z\}\le\underline l(s)$ and $\min\{\overline l(s),z\}\le\overline l(s)$. The monotonicity argument used in Step~2 of the proof of
Theorem~\ref{thm:soundness_monotone_bounds} then makes the sequences
$\{\underline V_1^{(k)}\}_{k\geq0}$ and
$\{\overline V_1^{(k)}\}_{k\geq0}$ non-increasing.
Each update selects among finitely many values, so both sequences converge after
finitely many iterations. We denote their limits by $\underline V_1^{\mathrm{cons}}$ and $\overline V_1^{\mathrm{cons}}$, which are fixed points of $\underline{\mathcal T}^{\mathrm{av}}$ and $\overline{\mathcal T}^{\mathrm{av}}$, respectively.

Proposition~\ref{prop:avoid} presents the avoid counterparts of Theorem~\ref{thm:soundness}   and of the guarantees of Proposition~\ref{prop:greedy}.

\begin{restatable}{proposition}{restateAvoid}
\label{prop:avoid}
Consider the avoid case with $\gamma=1$ and $\underline c\le\inf_{x\in\mathbb R^n}l(x)$, and fix some $x_0\in\mathcal X$ and $s_0\in S$ that satisfy $x_0\in[[s_0]]$. For each $s\in S$, let $\pi^*(s)\in\arg\max_{a\in A}\min_{s'\in\Delta(s,a)}\underline V_1^{\mathrm{cons}}(s')$.  Then,
\begin{enumerate}
\item $\underline V_1^{\mathrm{cons}}(s_0)\leq V_1(x_0)$. The trajectory of system~\eqref{eq:dynamics} starting from $x_0$ and following $u_\tau=\pi^*(s_\tau)$, where $s_\tau\in S$ satisfies $x_\tau\in[[s_\tau]]$, satisfies $l(x_\tau)\geq\underline V_1^{\mathrm{cons}}(s_0)$ for all $\tau\geq0$. If $\underline V_1^{\mathrm{cons}}(s_0)>0$, then $[[s_0]]$ does not intersect the BRS.
\item If $A=\mathcal U$, then $V_1(x_0)\leq\overline V_1^{(k)}(s_0)$ for all $k\geq0$. If, in addition, $\overline V_1^{(k)}(s_0)\leq0$ for some $k\geq0$, then, under every control sequence, every trajectory starting in $[[s_0]]$ reaches $\mathcal F$ at some time $\tau\leq k$, so $[[s_0]]$ is contained in the BRS.
\end{enumerate}
\end{restatable}
\begin{proof}
The proof is given in Appendix~\ref{app:avoid}.
\end{proof}

Consequently, since $[[s_\bot]]\subseteq\mathcal F$ and $\underline V_1^{\mathrm{cons}}(s_\bot)=\underline c\le0$, the union of the sets $[[s]]$, $s\in S$, with $\underline V_1^{\mathrm{cons}}(s)\le0$ is a sound over-approximation of the BRS, and, if $A=\mathcal U$, the union of the sets $[[s]]$ with $\overline V_1^{\mathrm{cons}}(s)  \leq  0$ is a sound under-approximation of the BRS.

\section{Case Studies}

\label{sec:case_studies}
We evaluate our method using five case studies with two- to five-dimensional state spaces: (1) Dubins car, (2) 3D evasion, (3) planar double integrator, (4) van der Pol oscillator, and (5) kinematic bicycle. We also empirically validate our theory in closed-loop simulations, including the soundness of the computed under-approximations of $\mathcal R$ and of the complement of the BRS, as well as the quantitative guarantees of Propositions~\ref{prop:greedy} and~\ref{prop:avoid}. In addition, we numerically validate the convergence bounds of Theorem~\ref{thm:convergence} on a variant of the Dubins car for which $f$ maps $\mathcal X\times\mathcal U$ into $\mathcal X$.

\subsection{Scenarios}
\label{sec:scenarios}

First, we describe the setup of the case studies.

\subsubsection{Dubins car}
We consider the following dynamics:
\begin{equation}
\label{eq:dubins}
\dot{x} = [v \cos x_3,\ v \sin x_3,\ u]^\top,
\end{equation}
where $[x_1, x_2]^\top \in [-3,3]^2$ represents the position of the vehicle in the plane, and $x_3 \in [-\pi, \pi]$ represents its heading angle. The vehicle moves with constant speed $v =1$, and the control input $u \in [-1, 1]$ specifies the angular velocity.
The failure function is defined as the signed distance to a cylindrical obstacle of radius $1.3$ centered at the origin: $l(x) = \sqrt{x_1^2 + x_2^2} - 1.3$.
The target set is defined as a circular region centered at $[2.5, 0]^\top$ with radius $0.5$. The reward function is defined as the negated signed distance to this target region: $r(x) = -\left(\sqrt{(x_1 - 2.5)^2 + x_2^2} - 0.5\right)$.

\subsubsection{3D evasion}
We consider a three-dimensional aircraft evasion scenario used in \cite{liu2025recurrentcontrolbarrierfunctions,mitchell2005HJ} with the following dynamics: $\dot{x} = [-v + v\cos x_3 + ux_2,\ v\sin x_3 - ux_1,\ -u]^\top$, where $[x_1, x_2]^\top \in [-3,3]^2$ denotes the relative planar position between the pursuer and evader, and $x_3 \in [-\pi, \pi]$ represents the relative heading angle. The pursuer moves with constant speed $v =1$, while the evader controls its angular velocity $u \in [-1, 1]$.
The failure function is defined as the signed distance to a cylindrical collision region of unit radius centered at the origin: $l(x) = \sqrt{x_1^2 + x_2^2} - 1$.
The target set and the reward function are defined as in the Dubins case.

\subsubsection{Planar double integrator}
We consider a planar double integrator with state \([x_1, x_2, x_3, x_4]^\top\), where \([x_1, x_2]^\top\) is the position and \([x_3, x_4]^\top\) is the velocity: \(\dot{x} = [x_3,\ x_4,\ u_1,\ u_2]^\top\).
The position lies in \([-3, 3]^2\) and the velocity in \([-1, 1]^2\). The control \(u = [u_1, u_2]^\top\) is the acceleration, with each coordinate taking values in the interval \([-1, 1]\). The failure function, the target set, and the reward function are the same as in the Dubins case.

\subsubsection{Van der Pol oscillator}
We consider the autonomous van der Pol oscillator with parameter \(\mu = 1\): \(\dot{x} = [x_2,\ (1 - x_1^2)\, x_2 - x_1]^\top\).
There is no control input. The state is restricted to the box \([-2.5, 2.5] \times [-2.9, 2.9]\).
The failure set is a disc of radius \(0.7\) centered at the origin: \(l(x) = \sqrt{x_1^2 + x_2^2} - 0.7\).
The target is a disc of radius \(0.4\) centered at \([2, 0]^\top\). The reward function is the negated signed distance to this disc: \(r(x) = -\left(\sqrt{(x_1 - 2)^2 + x_2^2} - 0.4\right)\).

\subsubsection{Kinematic bicycle}
We consider a kinematic bicycle with  state $[x_1, x_2, x_3, x_4, x_5]^\top$, where $[x_1, x_2]^\top$ is the position, $x_3 \in [-\pi, \pi]$ is the heading, $x_4$ is the speed, and $x_5$ is the steering angle: $\dot{x} = [x_4 \cos x_3,\ x_4 \sin x_3,\ x_4 \tan x_5,\ u_1,\ u_2]^\top$.
The position lies in $[-3, 3]^2$, the speed in $[-1, 1]$, and the steering angle in $[-0.6, 0.6]$.  The acceleration satisfies $u_1 \in [-0.2, 0.2]$, and the steering rate satisfies $u_2 \in [-1, 1]$. The failure function, the target set, and the reward function are the same as in the Dubins case.

\subsection{Results}

We construct discrete-time  dynamics by sampling the continuous-time ones using a  sampling time $t_s$. They are  
defined as  $x_{\tau+1} = f(x_\tau, a)$, which is  computed by numerically integrating the con\-tin\-u\-ous-time dynamics starting from $x_\tau$ over the time interval $[0, t_s]$ with the constant control signal that is equal to $a$. Validated ODE solvers, e.g., CAPD~\cite{kapela2021capd} and VNODE-LP~\cite{nedialkov2011vnodelp}, can bound integration error, which we can account for when over-approximating $R(s,a)$.
For all systems, $L_l=L_r=\sqrt2$ on $\mathcal X$, because $l$ and $r$ depend on $x$ in $\mathcal X$ only through Euclidean distances in the $(x_1,x_2)$-plane, which are $\sqrt2$-Lipschitz with respect to the infinity norm.  In these experiments, we restrict the admissible
control set $\mathcal U$ to the finite action set
$A$ specified below. Thus, $A=\mathcal U$, and the classifications of cells into $\mathcal R$ and $\mathcal L$ refer to these
finite-action systems.
For the Dubins car, we use 3 discrete actions $A = \{-1, 0, 1\}$,   
while for the 3D evasion task, we use 5 discrete actions $A = \{-1, -0.5, 0, 0.5, 1\}$.
For the planar double integrator, each acceleration component takes values in $\{-1, 0, 1\}$, so
$A = \{-1, 0, 1\}^2$ has nine actions.
The van der Pol oscillator is autonomous. For the kinematic bicycle, the acceleration takes values in $\{-0.2, 0, 0.2\}$ and the steering rate takes values in $\{-1, 0, 1\}$, so $A$ has nine actions.
For each system, $\mathcal X$ is the Cartesian product of the state-coordinate intervals specified in Section~\ref{sec:scenarios}. We set $l(x)=r(x)=-1$ for all $x\in\mathbb R^n\setminus\mathcal X$, which results in $\mathcal F\supseteq\mathbb R^n\setminus\mathcal X$, and we use $\underline c=\overline c=-1$, except in the Dubins avoid case, where $\underline c=-1.3=\min_{x\in\mathbb R^n}l(x)$ (Section~\ref{sec:avoid value function}). In Figures~\ref{fig:ourdub} and~\ref{fig:dubins_time}, iteration (iter) 1 is the computation on the initial grid, and each later iteration refines the grid as in Algorithm \ref{alg:alg2}.

\subsubsection{Analysis of the performance and the computational  requirements of Algorithm \ref{alg:alg2}}
\begin{figure}[ht]
\centering
\Description{Top: cell classifications for the Dubins car in the $(x_1,x_2)$ plane at several iterations of Algorithm 2. Bottom: runtime diagnostics of Algorithm 2 over its iterations.}
\begin{subfigure}{\linewidth}
    \centering
    \includegraphics[width=0.9\linewidth]{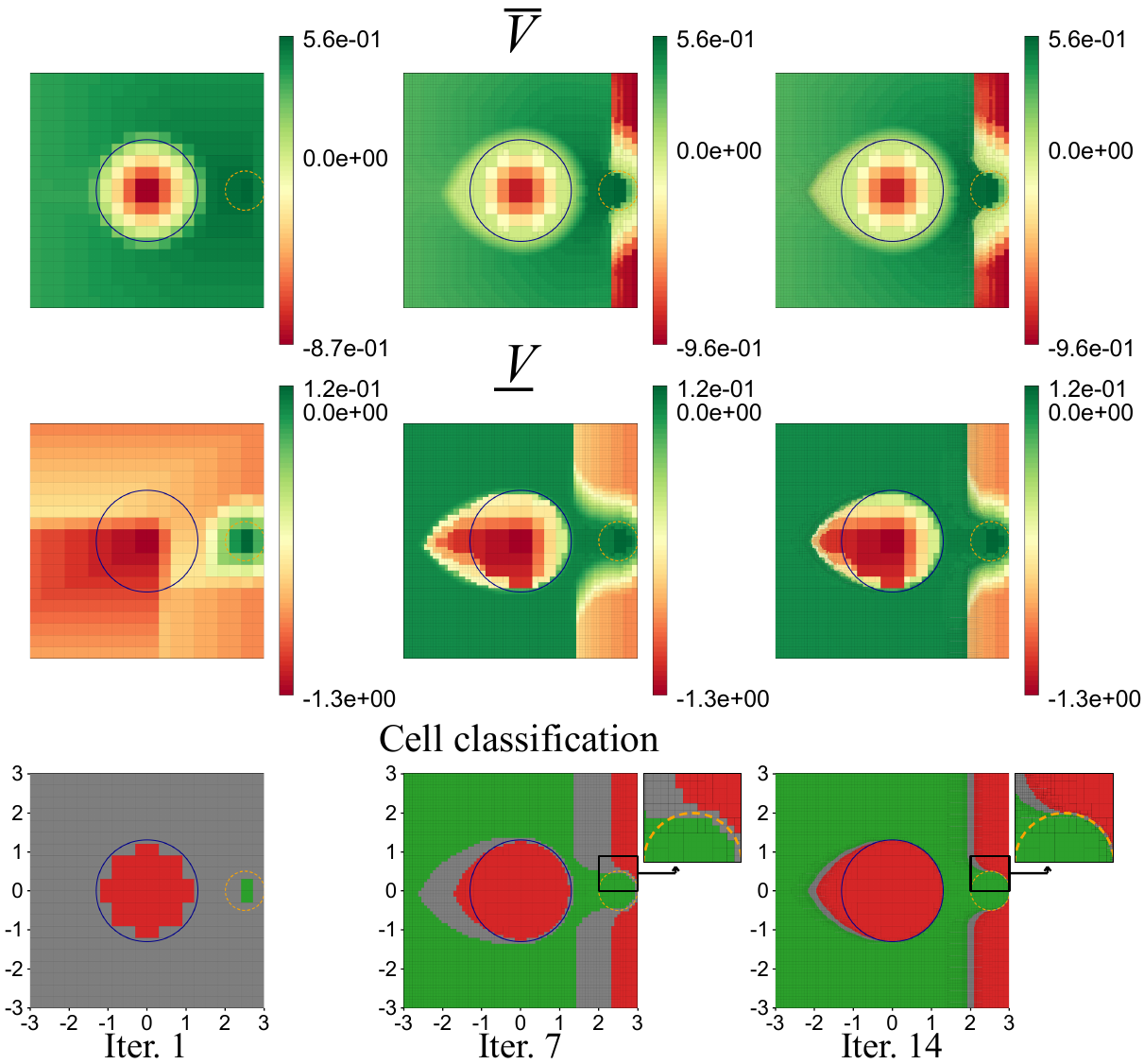}
    \caption{Output of Algorithm~\ref{alg:alg2} for the Dubins car with an initial uniform grid of $20^3$ cells, $t_s=0.3$, $\gamma = 0.96$, $\underline{\delta}^* = 10^{-8}$, $\overline{\delta}^* = 10^{-8}$, sliced at $\theta = 0$. The \textcolor{blue}{blue} circle defines $\mathcal{F}$. The \textcolor{mustard}{yellow} circle defines $\mathcal{T}$. Cells are \textcolor{darkgreen}{green} when $\underline V_\gamma^{\mathrm{cons}} > 0$, \textcolor{red}{red} when $\overline V_\gamma^{\mathrm{cons}} \leq 0$, and \textcolor{gray}{gray} otherwise.}
    \label{fig:ourdub}
\end{subfigure}

\vspace{0.5em}

\begin{subfigure}{1\linewidth}
    \centering
    \includegraphics[width=0.9\linewidth]{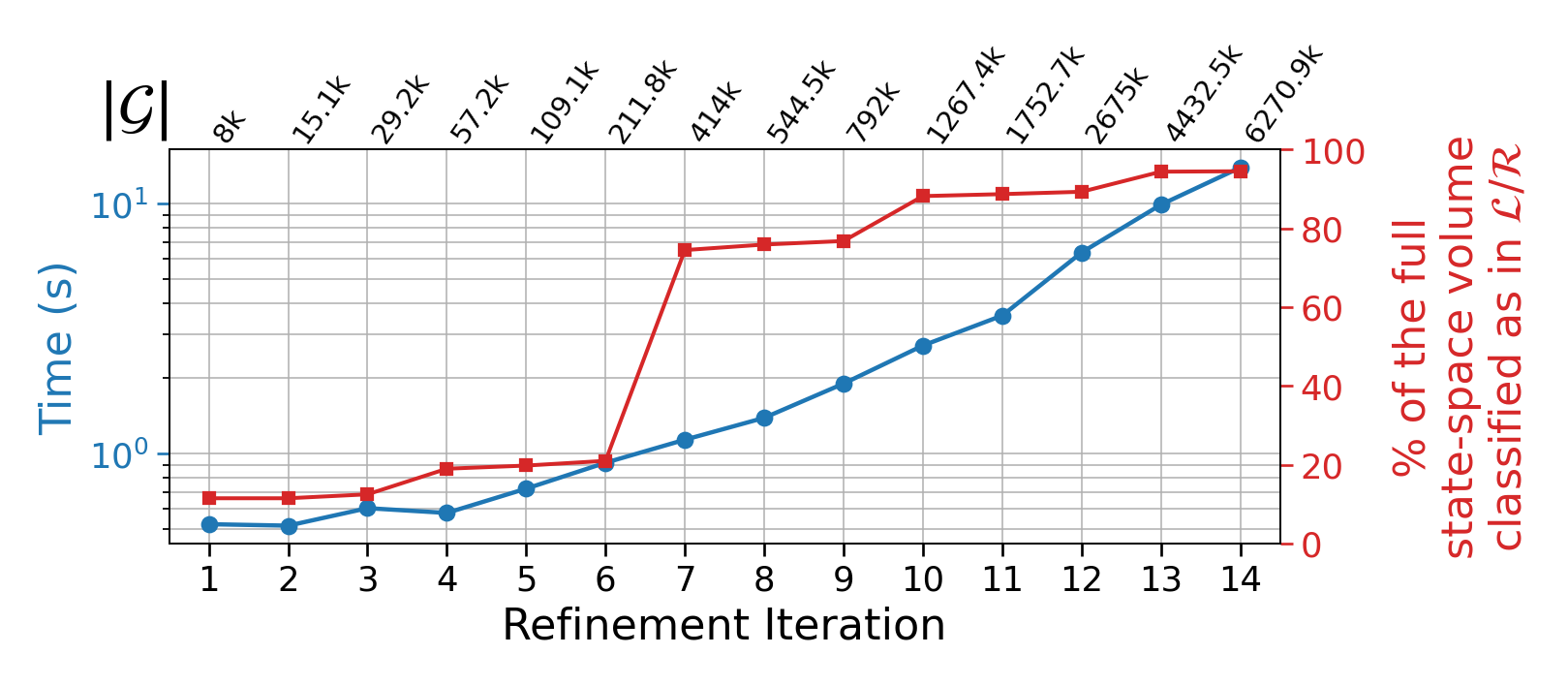}
    \caption{Runtime diagnostics of Algorithm \ref{alg:alg2}. }
    \label{fig:dubins_time}
\end{subfigure}
\caption{Dubins car: Algorithm~\ref{alg:alg2} output.}
\end{figure}

Figure \ref{fig:ourdub} shows the outputs of  Algorithm \ref{alg:alg2} at different iterations. Our method does not mislabel cells in $\mathcal{L}$ as being in $\mathcal{R}$ at any grid resolution, as guaranteed by Theorem \ref{thm:soundness}. As the number of refinement iterations increases, Algorithm~\ref{alg:alg2} certifies a larger fraction of the state-space volume as belonging to $\mathcal{R}$ or $\mathcal{L}$.

Figure \ref{fig:dubins_time} shows, for each iteration of Algorithm~\ref{alg:alg2} on the Dubins car, the runtime, the number of cells $|\mathcal G|$, and the percentage of the state-space volume certified as belonging to $\mathcal L$ or $\mathcal R$. Runtime accounts for the  unclassified cells' refinements, successor set computation, value iteration over the entire grid, and visualization, such as generating Figure \ref{fig:ourdub}. Experiments were conducted on a MacBook with an Apple M3 Pro chip and $18\,\mathrm{GB}$ of RAM.

\begin{figure}[h]
    \centering
    \includegraphics[width=0.9\columnwidth]{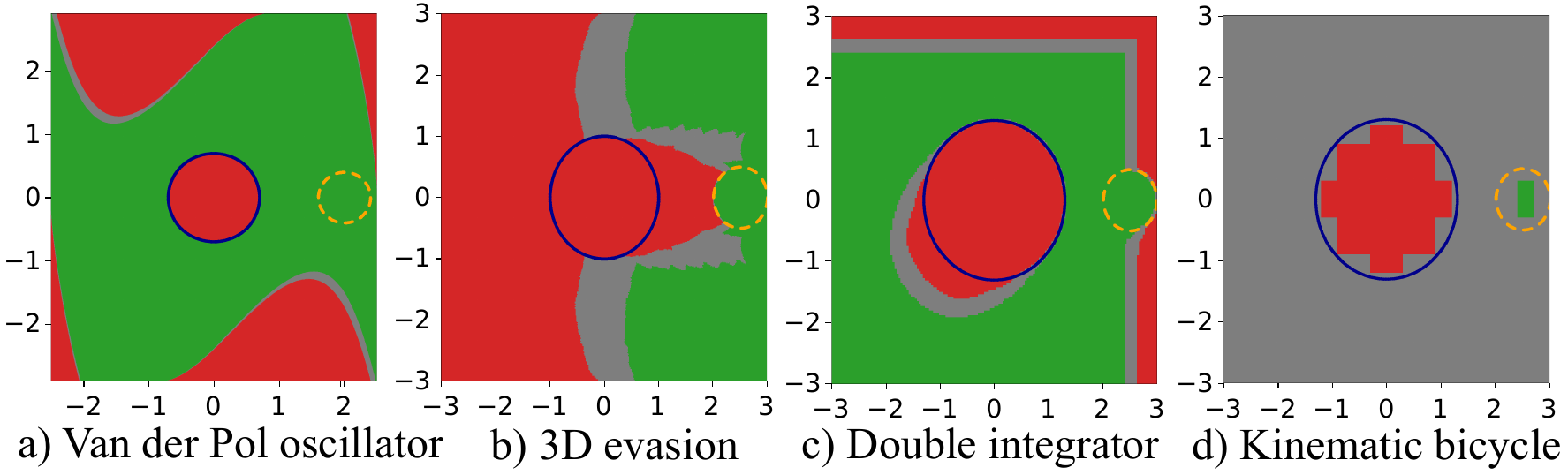}
    \Description{Cell classifications in the $(x_1,x_2)$ plane for the van der Pol oscillator, the 3D evasion model, the double integrator, and the kinematic bicycle. Green cells are certified in the reach-avoid set, red cells in its complement, and gray cells are unclassified.}
\caption{
    Cell classifications in the $(x_1,x_2)$ plane. Cells are \textcolor{darkgreen}{green} when $\underline V_\gamma^{\mathrm{cons}} > 0$, \textcolor{red}{red} when $\overline V_\gamma^{\mathrm{cons}} \leq 0$, and \textcolor{gray}{gray} otherwise. For each panel, we report the sampling time $t_s$, computation time, and the percentage
    of the state-space volume covered by cells certified
to lie in $\mathcal{R}$ or $\mathcal{L}$.
    (a) Van der Pol oscillator: full state space, 
    $t_s=0.01\,\mathrm{s}$, $291.75\,\mathrm{s}$,
    $97.7\%$ classified.
    (b) 3D evasion: slice at $\theta=\pi\approx3.14$,
    $t_s=0.3\,\mathrm{s}$, $1301.23\,\mathrm{s}$,
    $88.0\%$ classified.
    (c) Planar double integrator: slice at $(v_x,v_y)=(1,1)$,
    $t_s=0.3\,\mathrm{s}$, $1910.33\,\mathrm{s}$,
    $94.3\%$ classified.
    (d) Kinematic bicycle: slice at $(\theta,v,\delta)=(0,1,0)$, 
    $t_s=0.3\,\mathrm{s}$, $63.74\,\mathrm{s}$,
    $11.5\%$ classified.}
    \label{fig:ra_new_tasks}
\end{figure}

Figure~\ref{fig:ra_new_tasks} presents an evaluation of
Algorithm~\ref{alg:alg2} on systems with two through
five state dimensions. For the two- to four-dimensional systems, the algorithm classifies most of the full state space
as belonging to either $\mathcal L$ or $\mathcal R$ ($88.0$--$97.7\%$ of its volume), with runtime increasing with the state dimension. For the five-dimensional kinematic bicycle, Algorithm~\ref{alg:alg2} exceeded the available $18\,\mathrm{GB}$ of memory in the ninth refinement iteration, and Figure~\ref{fig:ra_new_tasks}(d) shows its output after eight iterations.
Our method inherits the scalability limitations of grid-based HJ reachability and symbolic control synthesis methods. Adaptive refinement reduces the need for uniformly fine grids, but storage for cells and their successor sets remains expensive. 

\subsubsection{Comparison between our method and existing tools}
We evaluate the performance of our proposed method against ROCS 2.0 \cite{ROCS2}, a formal control synthesis toolbox for nonlinear systems, and the existing HJ reachability toolbox \cite{mitchell2005toolbox}. 

ROCS 2.0 provides two synthesis engines: an abstraction-based engine that over-approximates system behaviors and a specification-guided engine that partitions the state space incrementally. In the specification-guided engine, $\varepsilon_{\min}$ defines the stopping criterion of an interval branch-and-bound refinement, as in our Algorithm~\ref{alg:alg2}. 

\begin{figure}[!t]
    \centering
    \includegraphics[width=0.9\linewidth]{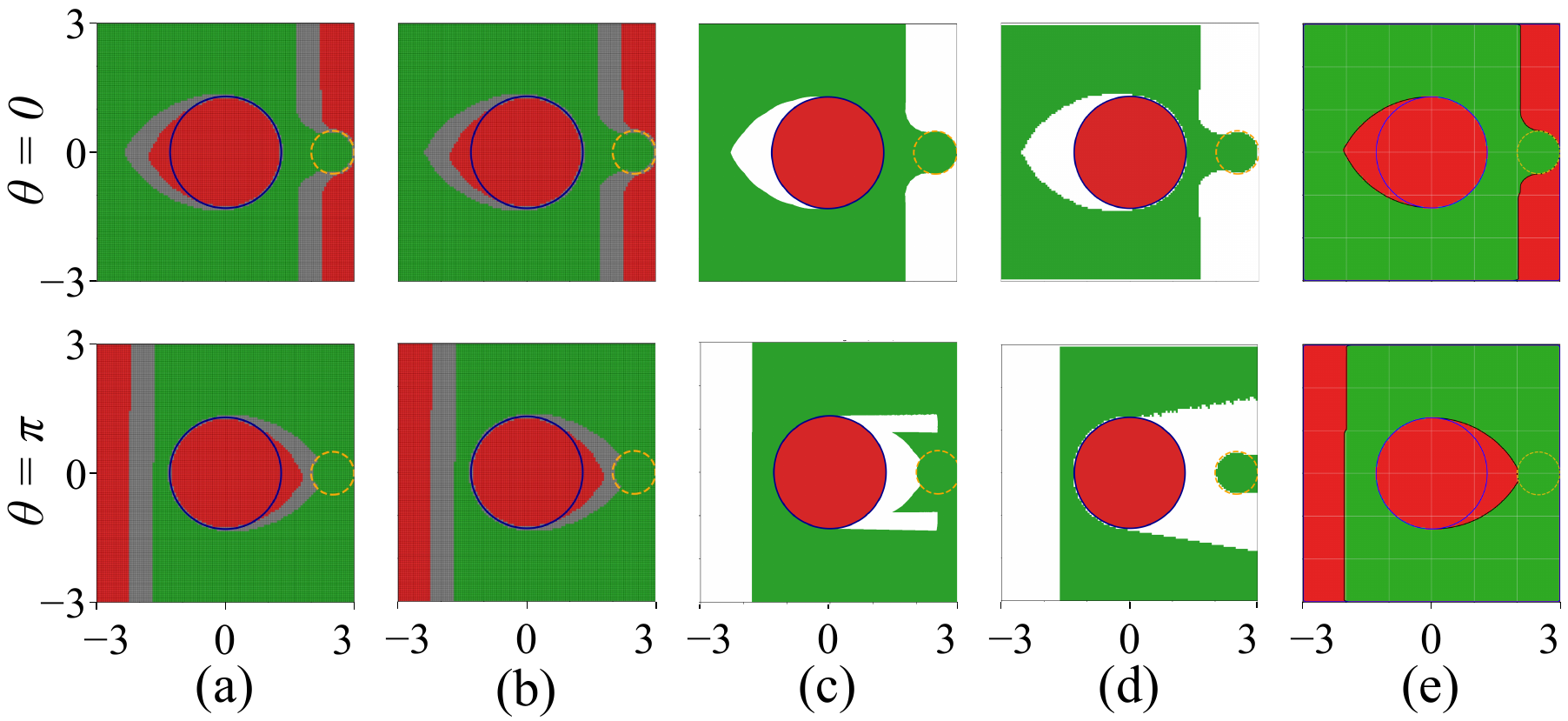}
    \Description{Reach-avoid sets for the Dubins car computed by Algorithm 1 with two discount factors, by ROCS 2.0 with two engines, and by an HJ PDE solver.}
    \caption{Reach-avoid set computations for the Dubins case study using different methods.  (a)  Algorithm~\ref{alg:alg1} with the modifications of Proposition~\ref{prop:gamma_1}, $120^3$ cells ($\varepsilon\approx0.026$), $t_s = 0.3$, $\gamma = 1$, running for $2.69\,\mathrm{s}$. (b) Algorithm~\ref{alg:alg1} with the same grid, $t_s = 0.3$, $\gamma = 0.96$, $\underline{\delta}^* = 10^{-8}$, $\overline{\delta}^* = 10^{-8}$, running for $2.86\,\mathrm{s}$. (c) ROCS 2.0 \cite{ROCS2} with $\varepsilon_{\min} = 0.005$, $t_s = 0.3$, running in $1443\,\mathrm{s}$. (d) ROCS 2.0 with $\varepsilon_{\min} = 0.026$ and $t_s = 0.3$, running for $18\,\mathrm{s}$. (e) HJ PDE solver \cite{mitchell2005toolbox} with $\varepsilon = 0.026$  and horizon $T=25\,\mathrm{s}$, running for $8046\,\mathrm{s}$. 
In (a)--(d), \textcolor{darkgreen}{green} and
\textcolor{red}{red} indicate certified membership in
$\mathcal R$ and $\mathcal L$, respectively, while
\textcolor{gray}{gray} or white indicates unclassified
states. In (e), they indicate the HJ PDE \cite{mitchell2005toolbox}   solver's numerical estimates, without guarantees.
    }

    \label{fig:related_works}
\end{figure}

Figures~\ref{fig:related_works}(a) and~\ref{fig:related_works}(b), where $\gamma=1$ and $\gamma=0.96$, respectively, show similar classifications.
The values of $\gamma$ in $(0.8,1]$ that we tested yield similar classifications when $\underline{\delta}^*,\overline{\delta}^*\leq10^{-8}$.
Larger tolerances can reduce runtime but certify smaller subsets of $\mathcal{L}$ and $\mathcal{R}$.

In our experiments, Algorithm~\ref{alg:alg1} certifies a larger part of the RAS than ROCS~2.0 in less time: it runs in $2.69\,\mathrm{s}$ and $2.86\,\mathrm{s}$ (Figures~\ref{fig:related_works}(a) and~\ref{fig:related_works}(b)), whereas ROCS~2.0 runs in $1443\,\mathrm{s}$ and $18\,\mathrm{s}$ (Figures~\ref{fig:related_works}(c) and~\ref{fig:related_works}(d)).
Our method also computes sound bounds on the reach-avoid value
function, which, in the discounted case, provide quantitative guarantees on safe target-reaching times, safety margins, and failure-reaching times (Proposition~\ref{prop:greedy}), whereas ROCS 2.0 provides binary classifications. Moreover, ROCS 2.0 certifies a subset of $\mathcal R$, but states outside this subset are not generally
certified to belong to $\mathcal L$. The HJ toolbox~\cite{mitchell2005toolbox}  solves HJ PDEs for continuous-time systems over a finite horizon, whereas our framework considers infinite-horizon reach-avoid problems for discrete-time systems. The sets that the toolbox computes at any grid resolution are not guaranteed to over- or under-approximate $\mathcal L$ and $\mathcal R$ (Figure~\ref{fig:related_works}(e)). 

\subsubsection{Avoid value function}
\label{sec:avoid_results}

\begin{figure}[ht]
\centering
\includegraphics[width=0.9\linewidth]{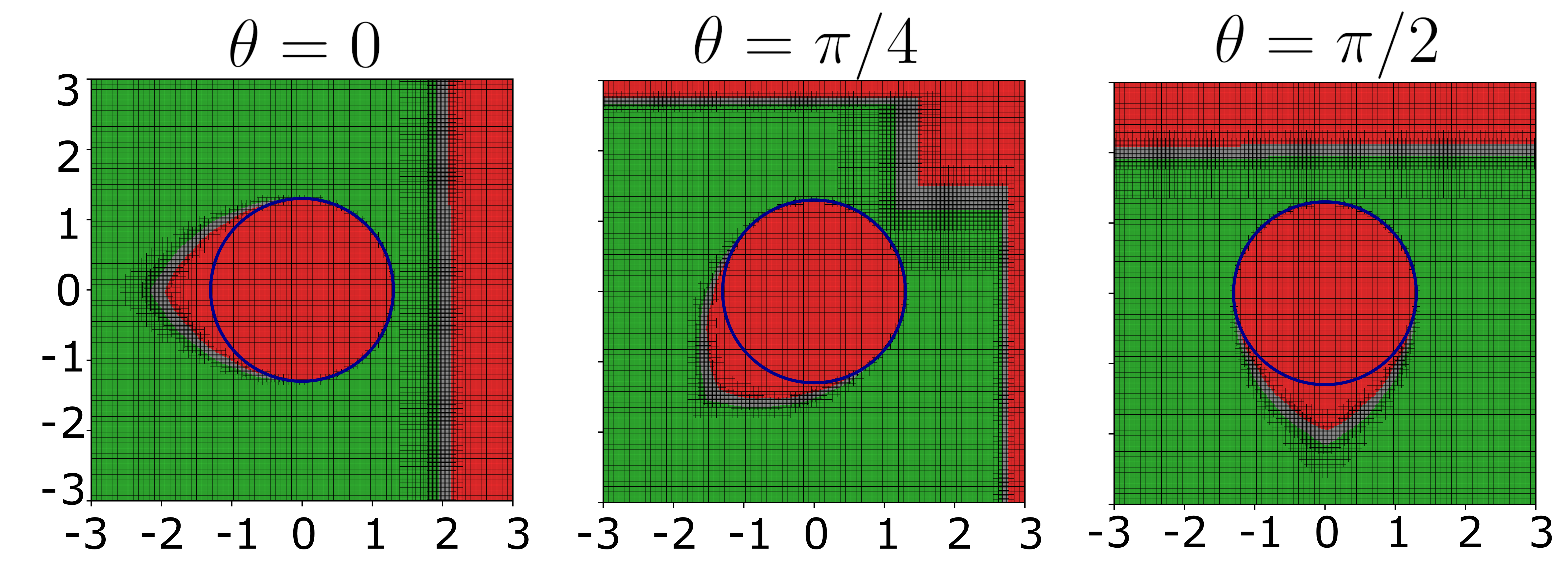}
\Description{Cell classifications for the Dubins car in the avoid case, computed by Algorithm 2. Green cells are certified outside the backward reachable set, red cells inside it, and gray cells are unclassified.}
\caption{Output of Algorithm~\ref{alg:alg2} with avoid operators, $t_s=0.3$, an initial uniform grid of $80^3$ cells, and six refinements ($22.1\,\mathrm{s}$): \textcolor{darkgreen}{green} when $\underline V_1^{\mathrm{cons}}>0$, \textcolor{red}{red} when $\overline V_1^{\mathrm{cons}}\leq0$, and \textcolor{gray}{gray} otherwise.}
\label{fig:avoid_dubins}
\end{figure}

Figure \ref{fig:avoid_dubins} shows the results of Algorithm \ref{alg:alg2} for the avoid case, i.e., when Algorithm~\ref{alg:alg1} uses the operators, initializations, and stopping criterion of Section~\ref{sec:avoid value function}.

\subsubsection{Empirical validation of the guarantees}
\label{sec:validation}
\emph{Reach-avoid case.} We run Algorithm~\ref{alg:alg1} for the Dubins car on a uniform grid of $120^3$ cells with $t_s=0.3$, $\gamma=0.96$, and $\underline\delta^*=\overline\delta^*=10^{-8}$. It terminates after $71$ iterations, with $\underline V_\gamma^{\mathrm{cons}}(s)>0$ in $983{,}988$ cells and $\overline V_\gamma^{\mathrm{cons}}(s)<0$ in $458{,}296$ cells. We simulated the system from one random state in each cell $s$ with $\underline V_\gamma^{\mathrm{cons}}(s)>0$ under $\hat\pi$. Every simulated trajectory reaches $\mathcal T$ at some time $t\leq\bar t:=\lfloor\log(\underline V_\gamma^{\mathrm{cons}}(s)/r^*)/\log(\gamma)\rfloor$ and satisfies $l(x_\tau)\geq\gamma^{-\tau}\underline V_\gamma^{\mathrm{cons}}(s)$ for $\tau=0,\ldots,t$, as Proposition~\ref{prop:greedy} guarantees. Also, for each cell $s$ with $\overline V_\gamma^{\mathrm{cons}}(s)<0$, we simulated the system from one random state in its interior and from its eight corners under every possible action sequence until $l(x_t)\leq\gamma^{-t}\overline V_\gamma^{\mathrm{cons}}(s)$. In every resulting trajectory,  the inequality was satisfied at some time $t\leq\bar t_{\mathcal F}:=\lfloor\log(\overline V_\gamma^{\mathrm{cons}}(s)/l_{\min})/\log(\gamma)\rfloor$, where $l_{\min}=-1.3$, with $r(x_\tau)\leq0$ for $\tau<t$, as the second claim of Proposition~\ref{prop:greedy} states.

\emph{Avoid case.} With the avoid operators, initializations, and stopping criterion of Section~\ref{sec:avoid value function}, Algorithm~\ref{alg:alg1} terminates on the same grid after $89$ iterations, with $\underline V_1^{\mathrm{cons}}(s)>0$ in $973{,}616$ cells and $\overline V_1^{\mathrm{cons}}(s)\leq0$ in $464{,}032$ cells. We simulated the system starting from one random state from each cell $s$ with $\underline V_1^{\mathrm{cons}}(s)>0$ while following $\pi^*$. Every resulting trajectory  satisfied $l(x_\tau)\geq\underline V_1^{\mathrm{cons}}(s)$ for $\tau=0,\ldots,1000$, as Proposition~\ref{prop:avoid} guarantees. As before, for each cell $s$ with $\overline V_1^{\mathrm{cons}}(s)\leq0$, we simulated the system starting from one random state in its interior and from its eight corners
while following every possible action sequence until it reaches $\mathcal F$. In every resulting trajectory, $\mathcal F$ was reached at some time $\tau\leq k$, where $k$ is the first iteration index satisfying $\overline V_1^{(k)}(s)\leq0$, as the second claim of Proposition~\ref{prop:avoid} states.

\emph{Convergence.} Appendix~\ref{app:bound_verification} empirically validates Theorem~\ref{thm:convergence}.

\section{Conclusion}
\label{sec:conclusion}
We presented a framework that uses value iteration on finite state-space grids to compute sound upper and lower bounds on avoid and reach-avoid value functions for discrete-time nonlinear control systems. It accounts for discretization errors and, for discounted reach-avoid problems, preserves sound bounds at every iteration and grid resolution. The obtained bounds provide quantitative guarantees on safe target-reaching times and safety margins in the discounted reach-avoid setting, on safety margins in the undiscounted avoid setting, and, when the control set is finite and $A=\mathcal U$, on failure-reaching times in both settings. Their signs also yield over-approximations of the BRS and under-approximations of the RAS, and, when the control set is finite and $A=\mathcal U$, sound under-approximations of the BRS and of the complement of the RAS.
We also presented a refinement algorithm that splits unclassified cells.
Future directions include GPU parallelization, local value-iteration updates~\cite{patching_2024_sylvia}, and extensions to continuous-time systems.
\bibliographystyle{abbrv}
\bibliography{ref}

\clearpage
\appendix
\makeatletter
\def\@appsecprefix@section{Appendix~}
\def\@seccntformat#1{\csname @appsecprefix@#1\endcsname\csname the#1\endcsname\quad}
\makeatother

\section{Proof of \texorpdfstring{Proposition~\ref{prop:discrete_contraction}}{Proposition}}
\label{app:contraction}
\restateContraction*

\begin{proof}
We prove (\ref{eq:contraction_upper}). The proof of (\ref{eq:contraction_lower}) is identical with $\max_{s'}$ replaced by $\min_{s'}$. Fix any $s \in S$. Let $M_i(s,a) := \max_{s' \in \Delta(s,a)} \gamma W_i(s')$, for $i=1,2$. Then, 

\begin{align*}
&\big|(\overline{\mathcal{T}}_\gamma W_1)(s) - (\overline{\mathcal{T}}_\gamma W_2)(s)\big| \\
&\le \Big|\max_{a} \max\{\overline{r}(s), M_1(s,a)\} - \max_{a} \max\{\overline{r}(s), M_2(s,a)\}\Big| \tag{i}\\
&\le \max_{a} \Big|\max\{\overline{r}(s), M_1(s,a)\} - \max\{\overline{r}(s), M_2(s,a)\}\Big| \tag{ii}\\
&\le \max_{a} |M_1(s,a) - M_2(s,a)| \tag{iii}\\
&\le \max_{a} \max_{s' \in \Delta(s,a)} |\gamma W_1(s') - \gamma W_2(s')| \tag{iv}\\
&= \gamma \max_{a} \max_{s' \in \Delta(s,a)} |W_1(s') - W_2(s')| \le \gamma \|W_1 - W_2\|_\infty. \tag{v}
\end{align*}
Taking the supremum over $s \in S$ yields~(\ref{eq:contraction_upper}). Step~(i) uses the identity $\max_a\min\{\overline l(s),y_a\}=\min\{\overline l(s),\max_a y_a\}$, which holds because $\overline l(s)$ does not depend on $a$, followed by the non-expansiveness of $\min$, i.e., $|\min\{a,b\}-\min\{c,d\}|\le\max\{|a-c|,|b-d|\}$. Step~(ii) uses the non-expansiveness of $\max$ over actions, $|\max_i a_i - \max_i b_i|\le \max_i|a_i-b_i|$. Step~(iii) uses $|\max\{c,a\}-\max\{c,b\}|\le|a-b|$. Step~(iv) uses the non-expansiveness of $\max$ over successors. Step~(v) uses the definition of $\|\cdot\|_\infty$.
\end{proof}

\section{Proof of \texorpdfstring{Proposition~\ref{prop:control_cells}}{Proposition}}
\label{app:control_cells}
\restateControlCells*

\begin{proof}
Let $s\in S\setminus\{s_\bot\}$ and $x\in[[s]]$ (for $s=s_\bot$, the claim holds as in Step~3 of the proof of Theorem~\ref{thm:soundness_monotone_bounds}), and let $u\in\mathcal U$ lie in the cell of the grid over $\mathcal U$ centered at $a\in A$, so that $\|u-a\|_\infty\leq\max_j\eta_{u,j}(a)$. By the triangle inequality and the Lipschitz continuity of $f$ in both arguments,
\begin{align*}
\|f(x,u)-f(x_c(s),a)\|_\infty
&\leq L_f\|x-x_c(s)\|_\infty+L_{f,u}\|u-a\|_\infty\\
&\leq L_f\max\nolimits_i\eta_{x,i}(s)+L_{f,u}\max\nolimits_j\eta_{u,j}(a),
\end{align*}
so $f(x,u)\in\bar R(s,a)$, and thus $f(x,u)\in[[s']]$ for some $s'\in\bar\Delta(s,a)$. Consequently, in the induction of Step~3 of the proof of Theorem~\ref{thm:soundness_monotone_bounds}, for every $u\in\mathcal U$ and the center $a\in A$ of a cell containing $u$, the induction hypothesis gives $V_\gamma(f(x,u))\leq\max_{s'\in\bar\Delta(s,a)}\overline V_\gamma^{(k)}(s')$, and therefore
\[
\min\{l(x),\max\{r(x),\gamma V_\gamma(f(x,u))\}\}\leq\overline V_\gamma^{(k+1)}(s).
\]
Taking the supremum over $u\in\mathcal U$ yields $V_\gamma(x)=\mathcal B_\gamma[V_\gamma](x)\leq\overline V_\gamma^{(k+1)}(s)$.
\end{proof}

\section{Proof of \texorpdfstring{Proposition~\ref{prop:gamma_1}}{Proposition}}
\label{app:gamma_1}
\restateGammaOne*

\begin{proof}
The initializations $\underline V_1^{(0)}=\min\{\underline l,\allowbreak\underline r\}$ and $\overline V_1^{(0)}=\min\{\overline l,\allowbreak\overline r\}$ equal the minimal and maximal RA measures~\eqref{eq:discrete_RA_measure_lower} and~\eqref{eq:discrete_RA_measure_upper} at $t=0$.
Both sequences are then non-decreasing by the argument used
in Step~2 of the proof of
Theorem~\ref{thm:soundness_monotone_bounds} to establish
$\underline V_\gamma^{(k)}\leq\underline V_\gamma^{(k+1)}$.
This argument remains valid when $\gamma=1$. 
Since $\gamma=1$, each update applies only minima and maxima to the cell-wise bounds and to the previous iterate, so every $\underline V_1^{(k)}$ takes values in the finite set $\{\underline l(s):s\in S\}\cup\{\underline r(s):s\in S\}$, and every $\overline V_1^{(k)}$ takes values in the finite set $\{\overline l(s):s\in S\}\cup\{\overline r(s):s\in S\}$. Thus, both
non-decreasing sequences converge after finitely many
iterations, at which point $\underline\delta_k=\overline\delta_k=0$ and Algorithm~\ref{alg:alg1} terminates.
The returned functions $\underline V_1^{\mathrm{cons}}$ and $\overline V_1^{\mathrm{cons}}$ are then fixed points of $\underline{\mathcal T}_1$ and $\overline{\mathcal T}_1$, respectively.

Define $W^{(0)}(x):=\min\{l(x),r(x)\}$ and
$W^{(k+1)}:=\mathcal B_1[W^{(k)}]$.
Unrolling this recursion shows that $W^{(k)}(x)$ is the supremum of $\max_{0\le t\le k}g_1(\xi,t)$ over the trajectories $\xi$ of~\eqref{eq:dynamics} with $x_0=x$ and inputs that may depend on time, where $g_1$ is $g_\gamma$ when $\gamma=1$.
Since the dynamics are deterministic and $\gamma=1$, repeated-state cycles can be removed from such a trajectory without decreasing its RA measure or increasing the time at which it is evaluated.
After removing all cycles up to the chosen evaluation time, the shortened finite prefix visits distinct states, so it can be generated by a stationary policy, extended arbitrarily elsewhere.
Thus, restricting the supremum to stationary policies does not change the finite-horizon value, i.e.,
$
W^{(k)}(x)
=\sup_{\pi}\max_{0\le t\le k}g_1(\xi_x^\pi,t)
$.
Consequently, $W^{(k)}(x)$ is non-decreasing in $k$ and,
by~\eqref{eq:Vgamma_def}, 
$\lim_{k\to\infty}W^{(k)}(x)=V_1(x)$. 

We prove by induction on $k$ that, for every $s\in S$
and $x\in[[s]]$,
$\underline V_1^{(k)}(s)\le W^{(k)}(x)$ if
$A\subseteq\mathcal U$, and
$W^{(k)}(x)\le\overline V_1^{(k)}(s)$ if
$A=\mathcal U$. \emph{Base case.}
At $k=0$, the cell-wise bounds give
$\underline V_1^{(0)}(s)
=\min\{\underline l(s),\underline r(s)\}$ \allowbreak
$\le W^{(0)}(x)=\min\{l(x),r(x)\}$ \allowbreak
$\le\min\{\overline l(s),\overline r(s)\}
=\overline V_1^{(0)}(s)$. \emph{Induction step.}
Assume the inequalities hold at iteration $k$ for every
$s\in S$ and $x\in[[s]]$. 
The argument follows Step~3 of the proof of
Theorem~\ref{thm:soundness_monotone_bounds}.
Here, $W^{(k)}$ replaces $V_\gamma$, and
$W^{(k+1)}=\mathcal B_1[W^{(k)}]$ replaces the fixed-point
equation $V_\gamma=\mathcal B_\gamma[V_\gamma]$. For $s=s_\bot$, the iterates equal $\underline c$ and $\overline c$, and every $x\in[[s_\bot]]$ satisfies $\underline c\le W^{(0)}(x)\le W^{(k)}(x)\le l(x)\le\overline c$. The bound
$\underline V_1^{(k+1)}(s)\le W^{(k+1)}(x)$
holds for $A\subseteq\mathcal U$ because maximizing over
$A$ cannot exceed taking the supremum over $\mathcal U$.
The bound
$W^{(k+1)}(x)\le\overline V_1^{(k+1)}(s)$
requires $A=\mathcal U$. Since $\{W^{(k)}(x)\}_{k\geq0}$ is non-decreasing with limit $V_1(x)$, this gives $\underline V_1^{(k)}(s)\le W^{(k)}(x)\le V_1(x)$ for every $k\geq0$ when $A\subseteq\mathcal U$, and, taking the limit, $V_1(x)\le\overline V_1^{\mathrm{cons}}(s)$ when $A=\mathcal U$.
Unlike for $\gamma<1$, an iterate $\overline V_1^{(k)}$ obtained before convergence need not upper-bound $V_1$, because it upper-bounds only $W^{(k)}$, which can be smaller than $V_1$. This is why the value iteration runs until $\overline\delta_k=0$.
The guarantees of Theorem~\ref{thm:soundness} therefore hold for $\underline V_1^{\mathrm{cons}}$ and $\overline V_1^{\mathrm{cons}}$, the fixed points returned by Algorithm~\ref{alg:alg1} when $\underline\delta_k=\overline\delta_k=0$.
\end{proof}

\section{Proof of \texorpdfstring{Proposition~\ref{prop:greedy}}{Proposition}}
\label{app:greedy}
\restateGreedy*

\begin{proof}
We first prove the claim for $\underline V_\gamma^{\mathrm{cons}}(s_0)>0$, which bounds the time to reach $\mathcal T$, and then the claim for $\overline V_\gamma^{\mathrm{cons}}(s_0)<0$, which bounds the time to reach $\mathcal F$.
Let $k$ be the iteration at which Algorithm~\ref{alg:alg1} terminates, so that $\underline V_\gamma^{\mathrm{cons}}=\underline V_\gamma^{(k)}$. For each $s\in S$, since $\underline l(s)$ and $\underline r(s)$ do not depend on $a$ and the minimum and maximum are non-decreasing in each argument, $\hat\pi(s)$ also attains the maximum over $a\in A$ in the definition of $(\underline{\mathcal T}_\gamma\underline V_\gamma^{\mathrm{cons}})(s)$. Moreover, $\underline V_\gamma^{(k+1)}\geq\underline V_\gamma^{(k)}$ by Corollary~\ref{cor:restricted_bounds}. Hence, for every $s\in S$,
\begin{align*}
&\min\Bigl\{\underline l(s),\max\Bigl\{\underline r(s),\gamma\min_{s'\in\Delta(s,\hat\pi(s))}\underline V_\gamma^{\mathrm{cons}}(s')\Bigr\}\Bigr\}\\
&\qquad=\underline V_\gamma^{(k+1)}(s)\geq\underline V_\gamma^{\mathrm{cons}}(s).
\end{align*}
Therefore, for every $s\in S$ and every $s'\in\Delta(s,\hat\pi(s))$,
\[
\underline V_\gamma^{\mathrm{cons}}(s)\leq\underline l(s)\quad\text{and}\quad\underline V_\gamma^{\mathrm{cons}}(s)\leq\max\bigl\{\underline r(s),\gamma\underline V_\gamma^{\mathrm{cons}}(s')\bigr\}.
\]
Let $\{x_\tau\}_{\tau\geq0}$ be the trajectory of system~\eqref{eq:dynamics} from $x_0$ under $u_\tau=\hat\pi(s_\tau)$, where $s_\tau\in S$ satisfies $x_\tau\in[[s_\tau]]$. Since $x_\tau\in[[s_\tau]]$ and $x_{\tau+1}=f(x_\tau,\hat\pi(s_\tau))\in[[s_{\tau+1}]]$, the construction of $\Delta$ in Section~\ref{sec:discrete_abstractions} gives $s_{\tau+1}\in\Delta(s_\tau,\hat\pi(s_\tau))$ for all $\tau\geq0$ with $s_\tau\neq s_\bot$. Let $t^*$ be the smallest $\tau\geq0$ with $\underline V_\gamma^{\mathrm{cons}}(s_\tau)\leq\underline r(s_\tau)$, with $t^*=\infty$ if no such $\tau$ exists. We show by induction on $\tau$ that
\[
\underline V_\gamma^{\mathrm{cons}}(s_\tau)\geq\gamma^{-\tau}\underline V_\gamma^{\mathrm{cons}}(s_0)\quad\text{for every }\tau\geq0\text{ with }\tau\leq t^*.
\]
\emph{Base case.} For $\tau=0$, both sides equal $\underline V_\gamma^{\mathrm{cons}}(s_0)$. \emph{Induction step.} Let $\tau<t^*$, and assume that $\underline V_\gamma^{\mathrm{cons}}(s_\tau)\geq\gamma^{-\tau}\underline V_\gamma^{\mathrm{cons}}(s_0)$. Then $\underline V_\gamma^{\mathrm{cons}}(s_\tau)>0\geq\underline c=\underline V_\gamma^{\mathrm{cons}}(s_\bot)$, so $s_\tau\neq s_\bot$. Since $\tau<t^*$, the definition of $t^*$ gives $\underline V_\gamma^{\mathrm{cons}}(s_\tau)>\underline r(s_\tau)$. Applying the inequality $\underline V_\gamma^{\mathrm{cons}}(s)\leq\max\{\underline r(s),\gamma\underline V_\gamma^{\mathrm{cons}}(s')\}$ with $s=s_\tau$ and $s'=s_{\tau+1}\in\Delta(s_\tau,\hat\pi(s_\tau))$ gives
\[
\underline V_\gamma^{\mathrm{cons}}(s_\tau)\leq\max\bigl\{\underline r(s_\tau),\gamma\underline V_\gamma^{\mathrm{cons}}(s_{\tau+1})\bigr\}.
\]
Since $\underline r(s_\tau)<\underline V_\gamma^{\mathrm{cons}}(s_\tau)$, this inequality holds only if $\underline V_\gamma^{\mathrm{cons}}(s_\tau)\leq\gamma\underline V_\gamma^{\mathrm{cons}}(s_{\tau+1})$. Therefore, by the induction hypothesis,
\[
\underline V_\gamma^{\mathrm{cons}}(s_{\tau+1})\geq\gamma^{-1}\underline V_\gamma^{\mathrm{cons}}(s_\tau)\geq\gamma^{-(\tau+1)}\underline V_\gamma^{\mathrm{cons}}(s_0),
\]
which completes the induction. If $t^*=\infty$, the induction gives $\underline V_\gamma^{\mathrm{cons}}(s_\tau)\geq\gamma^{-\tau}\underline V_\gamma^{\mathrm{cons}}(s_0)$ for every $\tau\geq0$. This is impossible: since $S$ is finite, $\underline V_\gamma^{\mathrm{cons}}$ takes finitely many values, whereas $\gamma^{-\tau}\underline V_\gamma^{\mathrm{cons}}(s_0)\to\infty$ as $\tau\to\infty$, because $\gamma\in(0,1)$ and $\underline V_\gamma^{\mathrm{cons}}(s_0)>0$. Hence, $t^*$ is finite. Let $\tau\leq t^*$. The cell-wise bound $l(x_\tau)\geq\underline l(s_\tau)$, the inequality $\underline V_\gamma^{\mathrm{cons}}(s_\tau)\leq\underline l(s_\tau)$, and the claim proved by induction give
\[
l(x_\tau)\geq\underline l(s_\tau)\geq\underline V_\gamma^{\mathrm{cons}}(s_\tau)\geq\gamma^{-\tau}\underline V_\gamma^{\mathrm{cons}}(s_0)>0,
\]
so $x_\tau\notin\mathcal F$. Moreover, the cell-wise bound $r(x_{t^*})\geq\underline r(s_{t^*})$, the inequality $\underline V_\gamma^{\mathrm{cons}}(s_{t^*})\leq\underline r(s_{t^*})$, which holds by the definition of $t^*$, and the claim proved by induction with $\tau=t^*$ give
\[
r(x_{t^*})\geq\underline r(s_{t^*})\geq\underline V_\gamma^{\mathrm{cons}}(s_{t^*})\geq\gamma^{-t^*}\underline V_\gamma^{\mathrm{cons}}(s_0)>0,
\]
so $x_{t^*}\in\mathcal T$. Thus, the trajectory reaches $\mathcal T$ at time $t^*$ and satisfies $l(x_\tau)\geq\gamma^{-\tau}\underline V_\gamma^{\mathrm{cons}}(s_0)>0$ for $\tau=0,\ldots,t^*$. Finally, $r^*\geq\underline r(s_{t^*})\geq\gamma^{-t^*}\underline V_\gamma^{\mathrm{cons}}(s_0)$ gives $\gamma^{t^*}\geq\underline V_\gamma^{\mathrm{cons}}(s_0)/r^*$. In particular, $r^*>0$ and $0<\underline V_\gamma^{\mathrm{cons}}(s_0)/r^*\leq\gamma^{t^*}\leq1$. Taking logarithms and dividing by $\log(\gamma)<0$, which reverses the inequality, gives $t^*\leq\log(\underline V_\gamma^{\mathrm{cons}}(s_0)/r^*)/\log(\gamma)$, and hence, since $t^*$ is an integer,
\[
t^*\leq\Bigl\lfloor\frac{\log(\underline V_\gamma^{\mathrm{cons}}(s_0)/r^*)}{\log(\gamma)}\Bigr\rfloor.
\]
This proves the first claim with $t=t^*$.

For the second claim, let $\kappa:=-\overline V_\gamma^{\mathrm{cons}}(s_0)>0$, let $\pi:\mathbb R^n\to\mathcal U$ be an arbitrary policy, and let $\{x_\tau\}_{\tau\geq0}=\xi_{x_0}^\pi$. Since $V_\gamma(x_0)$ in~\eqref{eq:Vgamma_def} is the supremum of the RA measure over all policies and times, $g_\gamma(\xi_{x_0}^\pi,\tau)\leq V_\gamma(x_0)$ for every $\tau\geq0$. Since $A=\mathcal U$, Corollary~\ref{cor:restricted_bounds} gives $V_\gamma(x_0)\leq\overline V_\gamma^{\mathrm{cons}}(s_0)=-\kappa$. Hence, $g_\gamma(\xi_{x_0}^\pi,\tau)\leq-\kappa$ for every $\tau\geq0$. Since $r$ is bounded and $\gamma\in(0,1)$, we have $\gamma^\tau r(x_\tau)>-\kappa$ for all sufficiently large $\tau$, and then~\eqref{eq:RA_measure} gives $\gamma^{\tau'}l(x_{\tau'})\leq-\kappa$ for some $\tau'\leq\tau$. Let $t$ be the smallest such $\tau'$. Then $l(x_t)\leq-\kappa\gamma^{-t}=\gamma^{-t}\overline V_\gamma^{\mathrm{cons}}(s_0)<0$, so $x_t\in\mathcal F$. Since $l_{\min}\leq l(x_t)\leq-\kappa\gamma^{-t}\leq-\kappa$, the existence of such a time implies $l_{\min}<0$ and $0<\kappa\leq|l_{\min}|$. Thus, the ratio $\overline V_\gamma^{\mathrm{cons}}(s_0)/l_{\min}=\kappa/|l_{\min}|$ lies in $(0,1]$, and the time bound in the statement is nonnegative. Moreover, $\gamma^tl_{\min}\leq\gamma^tl(x_t)\leq-\kappa$ gives $\gamma^t\geq \kappa/|l_{\min}|$. Taking logarithms and dividing by $\log(\gamma)<0$ gives $t\leq\log(\overline V_\gamma^{\mathrm{cons}}(s_0)/l_{\min})/\log(\gamma)$, and hence, since $t$ is an integer, the claimed bound. Finally, for $\tau<t$, the choice of $t$ gives $\gamma^{\tau'}l(x_{\tau'})>-\kappa$ for all $\tau'\leq\tau$, so $g_\gamma(\xi_{x_0}^\pi,\tau)\leq-\kappa$ forces $r(x_\tau)\leq\gamma^{-\tau}\overline V_\gamma^{\mathrm{cons}}(s_0)<0$, i.e., $x_\tau\notin\mathcal T$. Since also $x_t\in\mathcal F$ and $\mathcal T\cap\mathcal F=\emptyset$, the trajectory reaches $\mathcal F$ before $\mathcal T$, so it violates the RA specification. Since $\pi$ was arbitrary, this proves the second claim.
\end{proof}

\section{Proof of \texorpdfstring{Theorem~\ref{thm:convergence}}{Theorem}}
\label{app:convergence}
\restateConvergence*

\begin{proof}
We first show that $V_\gamma$ is $L$-Lipschitz on $\mathcal X$, following~\cite[Theorem~2]{li2025certifiable}, where $l$ plays the role of the constraint function $c$. That theorem considers the Euclidean norm, disturbances, and dynamics on $\mathbb R^n$. Let $Z:\mathbb R^n\to\mathbb R$ be bounded and $L$-Lipschitz on $\mathcal X$. For $x,y\in\mathcal X$ and $u\in\mathcal U$, we have $f(x,u),f(y,u)\in\mathcal X$ by assumption and $|\gamma Z(f(x,u))-\gamma Z(f(y,u))|\leq\gamma L_f L\|x-y\|_\infty\leq L\|x-y\|_\infty$, because $\gamma L_f<1$. Since $l$ and $r$ are $L$-Lipschitz on $\mathcal X$, and the minimum, the maximum, and the supremum over $u\in\mathcal U$ are non-expansive, $\mathcal B_\gamma[Z]$ is bounded and $L$-Lipschitz on $\mathcal X$. The function $\min\{l,r\}$ is bounded and $L$-Lipschitz on $\mathcal X$, and, since $\mathcal B_\gamma$ is a $\gamma$-contraction with fixed point $V_\gamma$ (Section~\ref{sec:HJ_value_function}), $\mathcal B_\gamma^k[\min\{l,r\}]$ converges uniformly to $V_\gamma$ as $k\to\infty$. A uniform limit of functions that are $L$-Lipschitz on $\mathcal X$ is $L$-Lipschitz on $\mathcal X$, so $V_\gamma$ is $L$-Lipschitz on $\mathcal X$. The rest of the proof has three steps. Step~1 samples $V_\gamma$ at the cell centers and shows that one application of $\overline{\mathcal T}_\gamma$ or $\underline{\mathcal T}_\gamma$ changes these samples by at most $(1+\gamma)L\varepsilon$. Step~2 uses the contraction property to show that the fixed points $\overline V_\gamma$ and $\underline V_\gamma$ lie within $(1+\gamma)L\varepsilon/(1-\gamma)$ of these samples. Step~3 extends this bound from the cell centers to all states of each cell. The last two paragraphs discuss the outputs of Algorithm~\ref{alg:alg1} and the unclassified cells.

\emph{Step 1: one-step error.} Since $f$ maps $\mathcal X\times\mathcal U$ into $\mathcal X$, $R(s,a)$ contains $f(x,a)$ for all $x\in[[s]]$, so it over-approximates the one-step reachable set as Section~\ref{sec:discrete_abstractions} requires, and the results of Sections~\ref{sec:valfunc_bounds}--\ref{sec:avoid value function} hold. Since $R(s,a)\subseteq\mathcal X$, no cell has $s_\bot$ as a successor, so removing $s_\bot$ from $S$ does not change the iterates and the fixed points on the cells. We therefore remove it in the rest of the proof. Define $W(s):=V_\gamma(x_c(s))$ for $s\in S$, fix $s\in S$, and write $x_c:=x_c(s)$. Since $V_\gamma=\mathcal B_\gamma[V_\gamma]$, $A=\mathcal U$, and the values $l(x_c)$, $r(x_c)$, $\overline l(s)$, and $\overline r(s)$ do not depend on $a$,
\[
W(s)=\min\Bigl\{l(x_c),\max\Bigl\{r(x_c),\gamma\max_{a\in A}V_\gamma(f(x_c,a))\Bigr\}\Bigr\},
\]
and
\[
(\overline{\mathcal T}_\gamma W)(s)=\min\Bigl\{\overline l(s),\max\Bigl\{\overline r(s),\gamma\max_{a\in A}\max_{s'\in\Delta(s,a)}W(s')\Bigr\}\Bigr\}.
\]
Let $a\in A$ and $s'\in\Delta(s,a)$. By the construction of $\Delta$ in Section~\ref{sec:discrete_abstractions}, with $R(s,a)=B_{L_f\varepsilon}(f(x_c,a))\cap\mathcal X$, there exists $y\in[[s']]\cap B_{L_f\varepsilon}(f(x_c,a))$. Since $\|x_c(s')-y\|_\infty\leq\varepsilon$ and $\|y-f(x_c,a)\|_\infty\leq L_f\varepsilon$, the triangle inequality gives $\|x_c(s')-f(x_c,a)\|_\infty\leq(L_f+1)\varepsilon$. The $L$-Lipschitz continuity of $V_\gamma$ then gives $|W(s')-V_\gamma(f(x_c,a))|\leq(L_f+1)L\varepsilon$. Moreover, $\Delta(s,a)$ is nonempty because a cell containing $f(x_c,a)\in\mathcal X$ belongs to it. Hence, the maximum and the minimum of $W$ over $\Delta(s,a)$ both differ from $V_\gamma(f(x_c,a))$ by at most $(L_f+1)L\varepsilon$. Since $\min$ and $\max$ are non-expansive, $|\overline l(s)-l(x_c)|\leq L\varepsilon$, and $|\overline r(s)-r(x_c)|\leq L\varepsilon$,
\[
|(\overline{\mathcal T}_\gamma W)(s)-W(s)|\leq\max\{L,\gamma(L_f+1)L\}\,\varepsilon\leq(1+\gamma)L\varepsilon,
\]
where $\max\{L,\gamma(L_f+1)L\}\leq(1+\gamma)L$ because $\gamma L_f<1$. The same argument with $\underline l$, $\underline r$, and $\min_{s'\in\Delta(s,a)}$ gives $|(\underline{\mathcal T}_\gamma W)(s)-W(s)|\leq(1+\gamma)L\varepsilon$.

\emph{Step 2: fixed points.} By Proposition~\ref{prop:discrete_contraction}, $\overline V_\gamma=\overline{\mathcal T}_\gamma\overline V_\gamma$ and $\overline{\mathcal T}_\gamma$ is a $\gamma$-contraction, so
\begin{align*}
\|\overline V_\gamma-W\|_\infty
&\leq\|\overline{\mathcal T}_\gamma\overline V_\gamma-\overline{\mathcal T}_\gamma W\|_\infty+\|\overline{\mathcal T}_\gamma W-W\|_\infty\\
&\leq\gamma\|\overline V_\gamma-W\|_\infty+(1+\gamma)L\varepsilon.
\end{align*}
Therefore, $\|\overline V_\gamma-W\|_\infty\leq(1+\gamma)L\varepsilon/(1-\gamma)$. The same argument gives $\|\underline V_\gamma-W\|_\infty\leq(1+\gamma)L\varepsilon/(1-\gamma)$.

\emph{Step 3: states inside a cell.} Let $x\in[[s]]$. Then $\|x-x_c(s)\|_\infty\leq\varepsilon$, so $|W(s)-V_\gamma(x)|\leq L\varepsilon$ because $V_\gamma$ is $L$-Lipschitz. Combining this with Step~2 through the triangle inequality gives
\[
|\overline V_\gamma(s)-V_\gamma(x)|\leq\frac{(1+\gamma)L\varepsilon}{1-\gamma}+L\varepsilon=\frac{2L\varepsilon}{1-\gamma},
\]
and the same bound holds for $\underline V_\gamma(s)$.

\emph{Outputs of Algorithm~\ref{alg:alg1}.} Let Algorithm~\ref{alg:alg1} terminate at iteration $k$. Since $\overline V_\gamma^{(k)}=\overline{\mathcal T}_\gamma\overline V_\gamma^{(k-1)}$ and $\overline{\mathcal T}_\gamma$ is a $\gamma$-contraction,
\begin{align*}
\|\overline V_\gamma^{(k)}-\overline V_\gamma\|_\infty
&\leq\gamma\|\overline V_\gamma^{(k-1)}-\overline V_\gamma\|_\infty\\
&\leq\gamma\bigl(\overline\delta_k+\|\overline V_\gamma^{(k)}-\overline V_\gamma\|_\infty\bigr),
\end{align*}
where $\overline\delta_k=\|\overline V_\gamma^{(k)}-\overline V_\gamma^{(k-1)}\|_\infty$. Rearranging and using $\overline\delta_k\leq\overline\delta^*$ and $\overline V_\gamma^{\mathrm{cons}}=\overline V_\gamma^{(k)}$ gives $\|\overline V_\gamma^{\mathrm{cons}}-\overline V_\gamma\|_\infty\leq\gamma\overline\delta^*/(1-\gamma)$. Similarly, $\|\underline V_\gamma^{\mathrm{cons}}-\underline V_\gamma\|_\infty\leq\gamma\underline\delta^*/(1-\gamma)$. Adding these bounds to those of Step~3 through the triangle inequality gives, for every $s\in S$ and $x\in[[s]]$, $|\overline V_\gamma^{\mathrm{cons}}(s)-V_\gamma(x)|\leq(2L\varepsilon+\gamma\overline\delta^*)/(1-\gamma)$ and $|\underline V_\gamma^{\mathrm{cons}}(s)-V_\gamma(x)|\leq(2L\varepsilon+\gamma\underline\delta^*)/(1-\gamma)$.

\emph{Unclassified cells.} If $\underline V_\gamma^{\mathrm{cons}}(s)\leq0<\overline V_\gamma^{\mathrm{cons}}(s)$ and $x\in[[s]]$, then the bounds on the outputs imply that $V_\gamma(x)\leq\underline V_\gamma^{\mathrm{cons}}(s)+(2L\varepsilon+\gamma\underline\delta^*)/(1-\gamma)\leq(2L\varepsilon+\gamma\underline\delta^*)/(1-\gamma)$ and $V_\gamma(x)\geq\overline V_\gamma^{\mathrm{cons}}(s)-(2L\varepsilon+\gamma\overline\delta^*)/(1-\gamma)>-(2L\varepsilon+\gamma\overline\delta^*)/(1-\gamma)$. Thus, $|V_\gamma(x)|\leq(2L\varepsilon+\gamma\max\{\underline\delta^*,\overline\delta^*\})/(1-\gamma)$.
\end{proof}

\section{Proof of \texorpdfstring{Proposition~\ref{prop:termination}}{Proposition}}
\label{app:termination}
\restateTermination*

\begin{proof}
For $0<\gamma<1$, contraction implies
$\underline\delta_k\to0$ and $\overline\delta_k\to0$.
Since $\underline\delta^*>0$ and $\overline\delta^*>0$,
Algorithm~\ref{alg:alg1} terminates in finite time.
For $\gamma=1$, Algorithm~\ref{alg:alg1} with the modifications of Proposition~\ref{prop:gamma_1} terminates after finitely many iterations by that proposition. With the avoid operators of Section~\ref{sec:avoid value function}, its iterates converge after finitely many iterations, as shown there, so it terminates as well.

In each iteration of Algorithm~\ref{alg:alg2}, all queued
cells are split along their longest axes. By construction
of the queue, each such axis has radius greater than
$\varepsilon_{\min}>0$, and each split halves that radius.
Thus, each axis can be split only finitely many times
along any branch of the refinement tree. The initial
grid is finite, so only finitely many splits are possible.
Since every iteration splits at least one cell, only
finitely many iterations are possible. Each iteration
then makes one call to Algorithm~\ref{alg:alg1} to
recompute the value function bounds over the refined grid.
Since each such call, as well as the initial call,
terminates, so does Algorithm~\ref{alg:alg2}.
\end{proof}

\emph{Cells omitted by the queue.} When $A\subset\mathcal U$, the queue of Algorithm~\ref{alg:alg2} omits the unclassified cells with $\overline V_\gamma^{\mathrm{cons}}(s)\le0$. Splitting them cannot certify them in $\mathcal R$. By the argument in the proof of Corollary~\ref{cor:restricted_bounds}, the value function $V_\gamma^A$ over policies with values in $A$ satisfies $\underline V_\gamma^{\mathrm{cons}}(s)\le V_\gamma^A(x)\le\overline V_\gamma^{\mathrm{cons}}(s)$ on every grid, for all $s\in S$ and $x\in[[s]]$. So $V_\gamma^A\le0$ on these cells, and $\underline V_\gamma^{\mathrm{cons}}\le0$ on every cell obtained by splitting them. The same argument applies for $\gamma=1$ and in the avoid case, with Propositions~\ref{prop:gamma_1} and~\ref{prop:avoid}, respectively, applied to the system with control set $A$ in place of Corollary~\ref{cor:restricted_bounds}.

\section{Proof of \texorpdfstring{Proposition~\ref{prop:avoid}}{Proposition}}
\label{app:avoid}
\restateAvoid*

\begin{proof}
We first show that both limits are attained after finitely many iterations. The operators $\underline{\mathcal T}^{\mathrm{av}}$ and $\overline{\mathcal T}^{\mathrm{av}}$ are monotone, since minima and maxima preserve pointwise inequalities. Since $\underline V_1^{(1)}=\underline{\mathcal T}^{\mathrm{av}}\underline l\leq\underline l=\underline V_1^{(0)}$, induction gives $\underline V_1^{(k+1)}\leq\underline V_1^{(k)}$ for all $k\geq0$, and the same argument applies to $\{\overline V_1^{(k)}\}_{k\geq0}$. Each update takes only minima and maxima of $\underline l$ and of the previous iterate, so every $\underline V_1^{(k)}$ takes values in the finite set $\{\underline l(s):s\in S\}$. Similarly, every $\overline V_1^{(k)}$ takes values in $\{\overline l(s):s\in S\}$. Since a non-increasing sequence in a finite set is eventually constant and $S$ is finite, there exists $K$ such that $\underline V_1^{\mathrm{cons}}=\underline V_1^{(K)}=\underline{\mathcal T}^{\mathrm{av}}\underline V_1^{(K)}$ and $\overline V_1^{\mathrm{cons}}=\overline V_1^{(K)}$.

(1) Recall that $\underline V_1^{\mathrm{cons}}=\underline{\mathcal T}^{\mathrm{av}}\underline V_1^{\mathrm{cons}}$ and that, for every $s\in S$,
\[
\pi^*(s)\in\arg\max_{a\in A}\min_{s'\in\Delta(s,a)}\underline V_1^{\mathrm{cons}}(s').
\]
By the definition of $\underline{\mathcal T}^{\mathrm{av}}$, it follows that, for every $s\in S$,
\[
\underline V_1^{\mathrm{cons}}(s)=\min\Bigl\{\underline l(s),\min_{s'\in\Delta(s,\pi^*(s))}\underline V_1^{\mathrm{cons}}(s')\Bigr\}.
\]
Hence, for all $s\in S$ and $s'\in\Delta(s,\pi^*(s))$, we have $\underline V_1^{\mathrm{cons}}(s)\leq\underline l(s)$ and $\underline V_1^{\mathrm{cons}}(s)\leq\underline V_1^{\mathrm{cons}}(s')$. The following argument applies to every choice of the states $s_\tau$. For the bound on $V_1(x_0)$, we take $s_\tau:=s_0$ if $x_\tau\in[[s_0]]$ and otherwise choose $s_\tau$ among the states $s\in S$ with $x_\tau\in[[s]]$ by a fixed rule, e.g., the one with the smallest index, so that $u_\tau$ depends only on $x_\tau$ and the choice at $\tau=0$ is $s_0$. Since $x_\tau\in[[s_\tau]]$ and $x_{\tau+1}=f(x_\tau,u_\tau)\in[[s_{\tau+1}]]$, where $u_\tau=\pi^*(s_\tau)$, the construction of $\Delta$ in Section~\ref{sec:discrete_abstractions} gives $s_{\tau+1}\in\Delta(s_\tau,u_\tau)$ whenever $s_\tau\neq s_\bot$, so $\underline V_1^{\mathrm{cons}}(s_{\tau+1})\geq\underline V_1^{\mathrm{cons}}(s_\tau)$. Let $\tau_1$ be the first time with $x_{\tau_1}\notin\mathcal X$, with $\tau_1=\infty$ if there is none. By induction on $\tau$, for all $\tau\leq\tau_1$,
\[
l(x_\tau)\geq\underline l(s_\tau)\geq\underline V_1^{\mathrm{cons}}(s_\tau)\geq\underline V_1^{\mathrm{cons}}(s_0).
\]
If $\tau_1<\infty$, then $s_{\tau_1}=s_\bot$, so $\underline V_1^{\mathrm{cons}}(s_0)\leq\underline V_1^{\mathrm{cons}}(s_\bot)=\underline c$, where the last equality holds because $\underline V_1^{(0)}(s_\bot)=\underline l(s_\bot)=\underline c$ and $(\underline{\mathcal T}^{\mathrm{av}}V)(s_\bot)=\min\{\underline c,V(s_\bot)\}$ for every $V$. After $\tau_1$, the trajectory may return to $\mathcal X$, but $\underline c\le\inf_{x\in\mathbb R^n}l(x)$ (Section~\ref{sec:avoid value function}) gives $l(x_\tau)\geq\underline c\geq\underline V_1^{\mathrm{cons}}(s_0)$ for all $\tau>\tau_1$ as well. Hence, $l(x_\tau)\geq\underline V_1^{\mathrm{cons}}(s_0)$ for all $\tau\geq0$. For this choice of the states $s_\tau$, the trajectory is generated by a policy that is admissible in~\eqref{eq:avoid_discounted_value_function}, so $V_1(x_0)\geq\inf_{\tau\geq0}l(x_\tau)\geq\underline V_1^{\mathrm{cons}}(s_0)$. If $\underline V_1^{\mathrm{cons}}(s_0)>0$, then $l(x_\tau)>0$, i.e., $x_\tau\notin\mathcal F$, for all $\tau\geq0$, so reaching $\mathcal F$ can be avoided from $x_0$, and $x_0$ is not in the BRS. Since $x_0\in[[s_0]]$ is arbitrary, $[[s_0]]$ does not intersect the BRS.

(2) Let $\pi$ be a policy, and let $\xi_{x_0}^\pi=\{x_\tau\}_{\tau\geq0}$ be the trajectory of system~\eqref{eq:dynamics} generated by $\pi$ from $x_0$. Then $\{x_{\tau+1}\}_{\tau\geq0}=\xi_{x_1}^\pi$ is the trajectory generated by $\pi$ from $x_1=f(x_0,\pi(x_0))$, so \eqref{eq:avoid_discounted_value_function} gives $\inf_{\tau\geq1}l(x_\tau)\leq V_1(x_1)$. Hence,
\begin{align*}
\inf_{\tau\geq0}l(x_\tau)&=\min\Bigl\{l(x_0),\inf_{\tau\geq1}l(x_\tau)\Bigr\}\\
&\leq\min\bigl\{l(x_0),V_1(f(x_0,\pi(x_0)))\bigr\},
\end{align*}
and taking the supremum over $\pi$ gives $V_1\leq\mathcal B_1^{\mathrm{av}}[V_1]$. We show by induction on $k$ that $V_1(x)\leq\overline V_1^{(k)}(s)$ for all $s\in S$ and $x\in[[s]]$. For $k=0$, $V_1(x)\leq l(x)\leq\overline l(s)=\overline V_1^{(0)}(s)$. Assume that the claim holds for some $k\geq0$. For $s=s_\bot$ and $x\in[[s_\bot]]$, $V_1(x)\leq l(x)\leq\overline c=\overline V_1^{(k+1)}(s_\bot)$. Now let $s\in S\setminus\{s_\bot\}$ and $x\in[[s]]$. For every $u\in\mathcal U=A$, there exists $s_u\in\Delta(s,u)$ with $f(x,u)\in[[s_u]]$, so $V_1(f(x,u))\leq\overline V_1^{(k)}(s_u)\leq\max_{s'\in\Delta(s,u)}\overline V_1^{(k)}(s')$. Therefore,
\begin{align*}
V_1(x)\leq\mathcal B_1^{\mathrm{av}}[V_1](x)
&\leq\min\Bigl\{\overline l(s),\max_{a\in A}\max_{s'\in\Delta(s,a)}\overline V_1^{(k)}(s')\Bigr\}\\
&=\overline V_1^{(k+1)}(s).
\end{align*}
Finally, let $\overline V_1^{(k)}(s_0)\leq0$ for some $k\geq0$. Suppose that some control sequence $u_0,u_1,\ldots\in\mathcal U$ generates, from some $y_0\in[[s_0]]$, a trajectory $\{y_\tau\}_{\tau\geq0}$ with $l(y_\tau)>0$ for $\tau=0,\ldots,k$. Since $\mathbb R^n\setminus\mathcal X\subseteq\mathcal F$, each such $y_\tau$ lies in $\mathcal X$, so $y_\tau\in[[s_\tau]]$ for some $s_\tau\in S\setminus\{s_\bot\}$, starting from the given $s_0$. Since $u_\tau\in\mathcal U=A$, the construction of $\Delta$ gives $s_{\tau+1}\in\Delta(s_\tau,u_\tau)$, and the definition of $\overline{\mathcal T}^{\mathrm{av}}$ gives $\overline V_1^{(j+1)}(s_\tau)\geq\min\{\overline l(s_\tau),\overline V_1^{(j)}(s_{\tau+1})\}$ for every $j\geq0$. Applying this inequality for $\tau=k-1,\ldots,0$ with $j=k-1-\tau$, starting from $\overline V_1^{(0)}(s_k)=\overline l(s_k)$, gives $\overline V_1^{(k)}(s_0)\geq\min_{\tau\leq k}\overline l(s_\tau)\geq\min_{\tau\leq k}l(y_\tau)>0$, a contradiction. Hence, under every control sequence, every trajectory starting in $[[s_0]]$ reaches $\mathcal F$ at some time $\tau\leq k$, and $[[s_0]]$ is contained in the BRS.
\end{proof}

\section{\texorpdfstring{Numerical validation of Theorem~\ref{thm:convergence}}{Numerical validation of Theorem}}
\label{app:bound_verification}

\begin{figure}[t]
    \centering
    \includegraphics[width=\columnwidth]{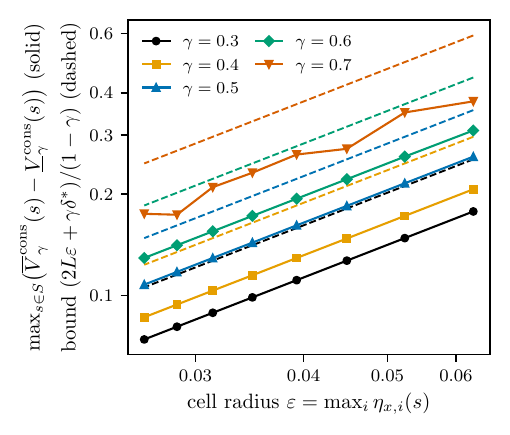}
    \Description{Largest gap between the computed upper and lower bounds, and the bound of Theorem 3, against the cell radius on logarithmic axes, for five discount factors.}
    \caption{$G=\max_{s\in S}(\overline V_\gamma^{\mathrm{cons}}(s)-\underline V_\gamma^{\mathrm{cons}}(s))$, computed from the outputs of Algorithm~\ref{alg:alg1} (solid), and the bound $(2L\varepsilon+\gamma\delta^*)/(1-\gamma)$ of Theorem~\ref{thm:convergence} (dashed) against the cell radius $\varepsilon$ for the periodic Dubins car, on logarithmic axes. Each color corresponds to one discount factor $\gamma$. Each solid curve below its dashed curve confirms the inequalities of Theorem~\ref{thm:convergence} for that $\gamma$.}
    \label{fig:bound_verification}
\end{figure}

This appendix verifies Theorem~\ref{thm:convergence} numerically on a periodic variant of the Dubins car of Section~\ref{sec:case_studies}. We validate its two claims on the outputs of Algorithm~\ref{alg:alg1}: (i) the bounds on the errors $|\overline V_\gamma^{\mathrm{cons}}(s)-V_\gamma(x)|$ and $|\underline V_\gamma^{\mathrm{cons}}(s)-V_\gamma(x)|$, and (ii) the bound on $|V_\gamma(x)|$ in the cells left unclassified. Since $V_\gamma$ is not known in closed form, we validate both claims through quantities that we can compute. Appendix~\ref{app:bv_setup} describes the system, its Lipschitz constants, and the runs of Algorithm~\ref{alg:alg1}, and Appendices~\ref{app:bv_errors} and~\ref{app:bv_unclassified} validate (i) and (ii), respectively. This validation covers only one system, five discount factors, and eight grids.

\subsection{\texorpdfstring{Setup}{Setup}}
\label{app:bv_setup}

\emph{Periodic Dubins car.} Theorem~\ref{thm:convergence} assumes that $f$ maps $\mathcal X\times\mathcal U$ into $\mathcal X$, and its proof uses the Lipschitz continuity of $V_\gamma$ on $\mathcal X$. In the experiments of Section~\ref{sec:case_studies}, the Dubins car can leave $\mathcal X$, where $l=r=-1$, so $V_\gamma$ can be discontinuous on $\mathcal X$, and Theorem~\ref{thm:convergence} does not apply. We therefore take $\mathcal X=[-3,3)^2\times[-\pi,\pi)$ and identify its opposite faces, i.e., we take $x_1$ and $x_2$ modulo $6$ and $x_3$ modulo $2\pi$. Then $\mathcal X$ is a three-dimensional torus, and $f$ maps $\mathcal X$ into $\mathcal X$, so the abstraction has no state $s_\bot$, and $S$ consists of the cells. We replace $\|x-y\|_\infty$ by the periodic distance
$
d(x,y):=\max_{i\in\{1,2,3\}}\min_{k\in\mathbb Z}|x_i-y_i-kp_i|,\quad p:=[6,6,2\pi]^\top,
$
in the definitions of the Lipschitz constants, of the cell radius, and of the balls $B_r(x)$. The results used below, namely Theorems~\ref{thm:soundness_monotone_bounds} and~\ref{thm:convergence}, use the norm only through the triangle inequality and the Lipschitz constants, so they hold with $d$ in place of $\|\cdot\|_\infty$. The experiments of Section~\ref{sec:case_studies} already take the heading $x_3$ of the Dubins car, the 3D evasion model, and the kinematic bicycle modulo $2\pi$, with the analogue of $d$ in which only $x_3$ is periodic. The results of Sections~\ref{sec:discrete_abstractions}--\ref{sec:avoid value function} hold with this distance for the same reason, and the Lipschitz constants of these systems remain valid for it, because their vector fields are $2\pi$-periodic in $x_3$ and $l$ and $r$ do not depend on $x_3$. In these experiments, a heading that passes $\pm\pi$ therefore does not leave $\mathcal X$, and $s_\bot\in\Delta(s,a)$ only if $R(s,a)$ leaves $\mathcal X$ in another coordinate. We use the dynamics~\eqref{eq:dubins} with $v=1$, $\mathcal U=A=\{-1,0,1\}$, and $t_s=0.3$, and we compute $f(x,u)$ exactly, as the closed-form solution of~\eqref{eq:dubins} at time $t_s$ from $x$ under the constant control $u$. The obstacle and the target are those of Section~\ref{sec:case_studies}, and $l$ and $r$ are the signed distances of Section~\ref{sec:case_studies} with the Euclidean distance in the plane replaced by the periodic Euclidean distance $d_2(z,z'):=\min_{k\in\mathbb Z^2}\|z-z'-6k\|_2$ for $z,z'\in\mathbb R^2$.

\emph{Lipschitz constants.} Since $d_2(z,z')\leq\sqrt2\max_{i\in\{1,2\}}\min_{k\in\mathbb Z}|z_i-z'_i-6k|$, the functions $l$ and $r$ are $\sqrt2$-Lipschitz with respect to $d$, so $L=\sqrt2$. The vector field in~\eqref{eq:dubins} is $1$-Lipschitz with respect to $\|\cdot\|_\infty$, since $|\cos a-\cos b|\leq|a-b|$ and $|\sin a-\sin b|\leq|a-b|$ for all $a,b\in\mathbb R$. By Gr\"onwall's inequality, the flow of~\eqref{eq:dubins} over $[0,t_s]$ under a constant control is $e^{t_s}$-Lipschitz with respect to $\|\cdot\|_\infty$ on $\mathbb R^3$. Since the vector field does not depend on $x_1$ and $x_2$ and is $2\pi$-periodic in $x_3$, the same holds with respect to $d$, so $L_f=e^{0.3}\approx1.350$. Thus, $\gamma L_f<1$ if and only if $\gamma<e^{-0.3}\approx0.741$, and we use $\gamma\in\{0.3,0.4,0.5,0.6,0.7\}$, for which $\gamma L_f$ ranges from $0.40$ to $0.94$.

\emph{Runs of Algorithm~\ref{alg:alg1}.} For each $\gamma$ and each $N\in\{50,60,\ldots,120\}$, we partition $\mathcal X$ into $N^3$ congruent cells with half-widths $3/N$, $3/N$, and $\pi/N$, so every cell has radius $\varepsilon=\pi/N$, which ranges from $0.026$ to $0.063$. We run Algorithm~\ref{alg:alg1} with $A=\mathcal U$, the cell bounds $\underline l$, $\overline l$, $\underline r$, and $\overline r$ of Section~\ref{sec:valfunc_bounds}, the transition relation $\Delta$ of Section~\ref{sec:discrete_abstractions} with $R(s,a)=B_{L_f\varepsilon}(f(x_c(s),a))$, and $\underline\delta^*=\overline\delta^*=\delta^*:=10^{-12}$. In all $40$ runs, it terminates after $24$ to $41$ iterations.

\subsection{\texorpdfstring{Error bounds on the outputs of Algorithm~\ref{alg:alg1}}{Error bounds on the outputs of Algorithm 1}}
\label{app:bv_errors}

\emph{Verified quantity.} With $\underline\delta^*=\overline\delta^*=\delta^*$, Theorem~\ref{thm:convergence} states that, for all $s\in S$ and $x\in[[s]]$,
\begin{align*}
|\overline V_\gamma^{\mathrm{cons}}(s)-V_\gamma(x)|&\leq(2L\varepsilon+\gamma\delta^*)/(1-\gamma),\\
|\underline V_\gamma^{\mathrm{cons}}(s)-V_\gamma(x)|&\leq(2L\varepsilon+\gamma\delta^*)/(1-\gamma).
\end{align*}
Since $V_\gamma$ is not known in closed form, these errors cannot be evaluated directly. Instead, we compute, from the outputs of Algorithm~\ref{alg:alg1}, the maximum over all cells of the difference between the upper and the lower bound,
\[
G:=\max_{s\in S}\bigl(\overline V_\gamma^{\mathrm{cons}}(s)-\underline V_\gamma^{\mathrm{cons}}(s)\bigr),
\]
and we check the inequality $G\leq(2L\varepsilon+\gamma\delta^*)/(1-\gamma)$, which is stronger than the two inequalities above. By Theorem~\ref{thm:soundness_monotone_bounds}, $\underline V_\gamma^{\mathrm{cons}}(s)\leq V_\gamma(x)\leq\overline V_\gamma^{\mathrm{cons}}(s)$ for all $s\in S$ and $x\in[[s]]$, so
\begin{align*}
|\overline V_\gamma^{\mathrm{cons}}(s)-V_\gamma(x)|&=\overline V_\gamma^{\mathrm{cons}}(s)-V_\gamma(x)\leq G,\\
|\underline V_\gamma^{\mathrm{cons}}(s)-V_\gamma(x)|&=V_\gamma(x)-\underline V_\gamma^{\mathrm{cons}}(s)\leq G.
\end{align*}
Thus, if $G\leq(2L\varepsilon+\gamma\delta^*)/(1-\gamma)$, then both inequalities of Theorem~\ref{thm:convergence} hold at every state of every cell.

\emph{Results.} Figure~\ref{fig:bound_verification} shows $G$ and the bound $(2L\varepsilon+\gamma\delta^*)/(1-\gamma)$ against $\varepsilon$. In all $40$ runs, $G$ lies below the bound, and $0.63\leq(1-\gamma)G/(2L\varepsilon+\gamma\delta^*)\leq0.73$. Thus, the inequalities of Theorem~\ref{thm:convergence} on the outputs of Algorithm~\ref{alg:alg1} hold at every state of every cell in all runs.

\subsection{\texorpdfstring{Bound on unclassified cells}{Bound on unclassified cells}}
\label{app:bv_unclassified}

We also check the last claim of Theorem~\ref{thm:convergence} on these $40$ runs. With $\underline\delta^*=\overline\delta^*=\delta^*$, this claim states that every cell $s$ with $\underline V_\gamma^{\mathrm{cons}}(s)\leq0<\overline V_\gamma^{\mathrm{cons}}(s)$ satisfies $|V_\gamma(x)|\leq\beta$ for all $x\in[[s]]$, where $\beta:=(2L\varepsilon+\gamma\delta^*)/(1-\gamma)$ is the error bound of Appendix~\ref{app:bv_errors}. Since $L=\sqrt2$ and $\delta^*=10^{-12}$, we have $\beta\approx2\sqrt2\,\varepsilon/(1-\gamma)$, which ranges from $0.106$ for $\gamma=0.3$ and $\varepsilon=0.026$ to $0.592$ for $\gamma=0.7$ and $\varepsilon=0.063$. We bound $|V_\gamma(x)|$ in two ways.

First, since $\underline V_\gamma^{\mathrm{cons}}(s)\leq V_\gamma(x)\leq\overline V_\gamma^{\mathrm{cons}}(s)$ by Theorem~\ref{thm:soundness_monotone_bounds}, every unclassified cell $s$ satisfies $|V_\gamma(x)|\leq M(s):=\max\{\overline V_\gamma^{\mathrm{cons}}(s),-\underline V_\gamma^{\mathrm{cons}}(s)\}$ for all $x\in[[s]]$. We evaluate $M(s)$ on all unclassified cells, between $14{,}756$ and $112{,}996$ per run. In all $40$ runs, the largest value of $M(s)/\beta$ lies between $0.58$ and $0.72$, so the claim holds at every state of every unclassified cell.

Second, we check the claim at sampled states with bounds on $V_\gamma$ that do not depend on Algorithm~\ref{alg:alg1}, whose outputs we use only to identify the unclassified cells. In each run, we pick $200$ unclassified cells uniformly at random without replacement, together with the two unclassified cells with the largest $\overline V_\gamma^{\mathrm{cons}}$ and the largest $-\underline V_\gamma^{\mathrm{cons}}$, and draw one state uniformly at random in each of these $202$ cells, with a fixed seed. At each such state $x$, we bound $V_\gamma(x)$ from both sides. We simulate, up to time $H$, the trajectories $\xi=\{x_t\}_{t\geq0}$ with $x_0=x$ generated by all $3^H$ action sequences $(u_0,\ldots,u_{H-1})\in A^H$, where $x_{t+1}=f(x_t,u_t)$ is the state reached from $x_t$ after time $t_s$ under~\eqref{eq:dubins} with the constant control $u_t$. Along each trajectory, we evaluate $g_\gamma(\xi,t)$ for $t=0,\ldots,H$, and we let $V_H(x)$ be the largest of these values over all trajectories and times. We now derive lower and upper bounds on $V_\gamma(x)$ from $V_H(x)$. For the lower bound, applying the Bellman equation $V_\gamma=\mathcal B_\gamma[V_\gamma]$ at $x_0,\ldots,x_t$ gives $V_\gamma(x)\geq g_\gamma(\xi,t)$ for every simulated trajectory $\xi$ and every $t\leq H$, so $V_\gamma(x)\geq V_H(x)$. For the upper bound, recall from~\eqref{eq:Vgamma_def} that $V_\gamma(x)$ is the supremum of $g_\gamma(\xi^\pi_x,t)$ over all policies $\pi$ and all times $t\geq0$. Each trajectory $\xi^\pi_x$ is generated by an action sequence, so up to time $H$ it coincides with the simulated trajectory of the first $H$ actions of this sequence. Hence, with $r_{\max}:=\max_{x\in\mathcal X}r(x)=0.5$, every value $g_\gamma(\xi^\pi_x,t)$ is at most $V_H(x)$ if $t\leq H$, and at most $\gamma^tr(x_t)\leq\gamma^{H+1}r_{\max}$ if $t>H$. Taking the supremum gives $V_\gamma(x)\leq\max\{V_H(x),\gamma^{H+1}r_{\max}\}$. Thus, for every $H\geq0$,
\[
V_H(x)\leq V_\gamma(x)\leq\max\{V_H(x),\gamma^{H+1}r_{\max}\}.
\]
Simulating only some action sequences would still result in the lower bound, but not the upper bound, which requires the search to be exhaustive.

We check the two conditions
\[
V_H(x)\geq-\beta\qquad\text{and}\qquad\max\{V_H(x),\gamma^{H+1}r_{\max}\}\leq\beta.
\]
If both hold, then
\[
-\beta\leq V_H(x)\leq V_\gamma(x)\leq\max\{V_H(x),\gamma^{H+1}r_{\max}\}\leq\beta,
\]
where the outer inequalities are the two conditions and the inner ones are the bounds derived above. Since the two bounds hold for every $H$, no choice of $H$ lets this check pass at a state where the claim fails: if $V_\gamma(x)<-\beta$, then $V_H(x)<-\beta$, and if $V_\gamma(x)>\beta$, then $\max\{V_H(x),\gamma^{H+1}r_{\max}\}>\beta$. The horizon affects only how tight the bounds are, and hence whether the check is conclusive. Increasing $H$ can only raise $V_H(x)$, which tends to $V_\gamma(x)$, and lowers $\gamma^{H+1}r_{\max}$, but the search grows as $3^H$. Decreasing $H$ can leave the claim unconfirmed at some states, and at every state once $\gamma^{H+1}r_{\max}>\beta$, since the upper bound then exceeds $\beta$. We choose $H$ by a heuristic rule that depends only on $\gamma$: the smallest $H$ with $\gamma^{H+1}r_{\max}\leq0.01$, namely $H=3$, $4$, $5$, $7$, and $10$ for $\gamma=0.3$, $0.4$, $0.5$, $0.6$, and $0.7$, for which the search covers at most $3^{10}=59{,}049$ action sequences per state.  Since $\beta$ ranges from $0.106$ to $0.592$, we have $\gamma^{H+1}r_{\max}\leq0.01<\beta/10$ in every run. Hence, the second condition, $\max\{V_H(x),\gamma^{H+1}r_{\max}\}\leq\beta$, holds if and only if $V_H(x)\leq\beta$. Together with the first condition, $V_H(x)\geq-\beta$, the two conditions are thus equivalent to $-\beta\leq V_H(x)\leq\beta$, that is, $|V_H(x)|\leq\beta$.

Both conditions hold at all $8{,}080$ sampled states, $202$ in each of the $40$ runs. Since the two bounds enclose $V_\gamma(x)$, this confirms both $V_\gamma(x)\geq-\beta$ and $V_\gamma(x)\leq\beta$. The two bounds also imply that $|V_\gamma(x)|\leq\max\{|V_H(x)|,\gamma^{H+1}r_{\max}\}$, and in each run, the largest value of the right-hand side over the $202$ states lies between $0.29\,\beta$ and $0.70\,\beta$. Thus, $|V_\gamma(x)|\leq0.70\,\beta$ at every sampled state.

\end{document}